\documentclass[11pt,preprint2]{aastex}
\usepackage{graphicx}
\usepackage{natbib}
\usepackage{lscape}

\shorttitle{Hydrogen permitted lines in the Th\,28 microjet}
\shortauthors{Coffey~et~al.(2010}

\begin{document}

\title{Hydrogen permitted lines in the first near-IR spectra of Th 28 microjet: accretion or ejection tracers?}

\author{
Deirdre Coffey\altaffilmark{1}, Francesca Bacciotti\altaffilmark{2}, 
Linda Podio\altaffilmark{1}, Brunella Nisini\altaffilmark{2}}

\altaffiltext{1}{The Dublin Institute for Advanced Studies, Dublin 2, Ireland \email{dac@cp.dias.ie}}
\altaffiltext{2}{Osservatorio Astrofisico di Arcetri, Largo E. Fermi 5, 50125 Firenze, Italy \email{fran@arcetri.astro.it}}
\altaffiltext{3}{Osservatorio di Roma, Frascati, Roma, Italy \email{nisini@oa-roma.inaf.it}
}

\begin{abstract}

\noindent
We report the first near-infrared detection of the bipolar microjet from T Tauri star ThA 15-28 (hereafter Th 28). Spectra were obtained with Very Large Telescope(VLT)/ISAAC for the slit both perpendicular and parallel to the flow to examine jet kinematics and gas physics within the first arcsecond from the star. The jet was successfully detected in both molecular and atomic lines. The H$_2$ component was found to be entirely blueshifted around the base of the bipolar jet. It shows that only the blue lobe is emitting in H$_2$ while light is scattered in the direction of the red lobe, highlighting an asymmetric extinction and/or excitation between the two lobes. Consistent with this view, the red lobe is brighter in all atomic lines. Interestingly, the jet was detected not only in [\ion{Fe}{2}], but also in Br$\gamma$ and Pa$\beta$ lines. Though considered tracers mainly of accretion, we find that these high excitation hydrogen permitted lines trace the jet as far as 150 AU from the star. This is confirmed in a number of ways: the presence of the [\ion{Fe}{2}] 2.13 $\micron$ line which is of similarly high excitation; \ion{H}{1} velocities which match the jet [\ion{Fe}{2}] velocities in both the blue and red lobe; and high electron density close to the source of $>$6$\times$10$^{4}$~cm$^{-3}$ derived from the [\ion{Fe}{2}] 1.64,1.60 $\micron$ ratio. These near-infrared data complement {\it Hubble Space Telescope} Imaging Spectrograph ({\it HST}/STIS) optical and near-ultraviolet data for the same target which were used in a jet rotation study, although no rotation signature could be identified here due to insufficient angular resolution. The unpublished {\it HST}/STIS H$\alpha$ emission is included here along side the other \ion{H}{1} lines. Identifying Br$\gamma$ and Pa$\beta$ as tracers of ejection is significant because of the importance of finding strong near-infrared probes close to the star, where forbidden lines are quenched, which will help understand accretion-ejection when observed with high spatial resolution instruments such as VLTI/AMBER. 

\end{abstract}


\section{Introduction}
\label{introduction}

Both observational and theoretical studies of the star formation process have unveiled a complex picture of simultaneous accretion and ejection which is yet to be fully understood. It is clear, however, that there exists a tight correlation between the two mechanisms, as predicted by theoretical models (e.g., \citealp{Konigl00}) and supported by observations (e.g., \citealp{Hartigan95}). 

Explaining the underlying physics rests ultimately on identifying the origin of observed emission of different atomic species, in order to determine the gas properties, structure and dynamics. For example, the diffuse gas in protostellar outflows produces numerous permitted and forbidden lines, whose excitation properties are reasonably well known. Observations of forbidden emission lines (such as O$^{0}$, N$^{+}$ and S$^{+}$ at optical wavelengths and Fe$^{+}$ in near-infrared (near-IR) wavelengths) show that they trace high velocity, low density outflow activity (e.g., \citealp{Podio06}). Permitted hydrogen lines, however, are more complex to analyse and so their exact origin is unclear. Advancements in understanding often rely on correlations and modeling rather than direct observation, particularly concerning the sub-arcsecond spatial scale on which the accretion/ejection engine operates. Hydrogen line profiles have been observed and modeled extensively in the literature (see \citealp{Reipurth96}; \citealp{Alencar00}; \citealp{Folha01} and references therein). Initial claims determined Balmer emission such as H$\alpha$ to be attributed to outflowing material (e.g., \citealp{Hartmann90}) while, subsequently, magnetospheric accretions models have investigated their origin in terms of infall (e.g., \citealp{Alencar00}). Indeed, the higher excitation Pa$\beta$ and Br$\gamma$ lines have been found to correlate highly with accretion luminosity in classical T Tauri stars and brown dwarfs (\citealp{Muzerolle98a} 1998b; \citealp{Natta04}). As a result, they have been used as mass accretion flux indicators in cases of highly embedded Class 0/1 sources where extinction is high, thus rendering the UV excess flux (i.e., the usual accretion indicator) difficult to measure accurately (e.g., \citealp{Garcialopez06}). However, there has never been satisfactory agreement between models and observations, or full consensus within the community as to the true origin of these lines. More recent studies have again suggested an origin in outflowing material. There is some evidence that Pa$\beta$ seems to trace T Tauri jet activity (\citealp{Takami01}; \citealp{Whelan04}; 2009). Also, interferometric observations imply the Br$\gamma$ line of Herbig Ae/Be stars may come from the inner disk region and/or an ejected wind \citep{Kraus08}. Meanwhile, Br$\gamma$ detections within 100\,AU of T Tauri stars have been suggested as originating in the outflow or scattered from an outflow cavity \cite{Beck09}. 

In this paper, we examine the bipolar jet for T Tauri star ThA\,15-28 (hereafter Th 28) near the jet base ($<$1$\arcsec$ along the jet, i.e., within 170\,AU of the star). This is close to the region where the plasma is accelerated and collimated (i.e., within a few tens of AU of the star) just after jet launching takes place. The original paper on this star was by \cite{Krautter86}, hence Th\,28 is now also known as Krautter's star. It is a classical T\,Tauri star located in the Lupus\,3 cloud (Th{\'e} 1962), which is at a distance of 170\,pc \cite{Eggen83}, and is surrounded in nebulosity (\citealp{Graham88}). The star's systemic heliocentric velocity, measured from photospheric lines, was found to be +5\,km\,s$^{-1}$ (\citealp{Graham88}). The outflow inclination angle with respect to the plane of the sky was measured as 10\degr, and the jet position angle is 98\degr (\citealp{Krautter86}). A study of the gas physics, based on ratios of optical forbidden emission lines, at 2$\arcsec$ along the flow showed the receding jet to be of high ionisation (i.e., $x_e$$\sim$0.5) \cite{BE99}. Recent optical images confirm high proper motion of the bipolar flow HH\,228 confirming that it is excited by Th 28, see Figure\,\ref{wang} \cite{Wang09}. 

The fact that this system is close to edge-on makes it ideal for examining the initial portion of the jet. 
Indeed, this bipolar jet was included in a survey on T Tauri jet targets which reported indications of jet rotation based on {\it Hubble Space Telescope} Imaging Spectrograph ({\it HST}/STIS) spectra at optical and ultraviolet wavelengths (\citealp{Coffey04}; 2007). The jet base was also the subject of a diagnostic study of the gas physics, which accessed jet plasma conditions within the first arcsecond from the star at high spatial resolution \cite{Coffey08}. These studies had important implications for the role of jets as angular momentum extractors, allowing accretion to proceed until the star reaches its final pre-main sequence mass. 

To build a multi-wavelength picture of this bipolar jet close to the source, we now choose to examine the Th 28 jet in near-IR emission. The near-IR is expected to access the atomic jet via [\ion{Fe}{2}] lines and the molecular flow via H$_2$ emission. The aim of the observations was to understand the physics that can be extracted from the diagnostics of the infrared lines. 
For example, the electron density of the jet close to the star usually saturates the optical [\ion{S}{2}]6731/6716~\AA~ratio, whereas the near-IR allows access to higher electron densities before saturation of the [\ion{Fe}{2}]1.64/1.60 $\micron$ ratio, since [\ion{Fe}{2}] has a higher critical density ($\sim$6$\times$10$^{4}$\,cm$^{-3}$) than [\ion{S}{2}] ($\sim$2$\times$10$^{4}$\,cm$^{-3}$). Higher densities determined in the initial jet channel have implications for mass and momentum fluxes, and hence the underlying mass loading at the jet base. Second, it is important to understand the relationship between the atomic and molecular emission, both of which are accessed via near-IR observations. Finally, observations with the perpendicular slit were intended to determine if rotation signatures are confirmed for the atomic and molecular flow in the near-IR, as have been reported in optical lines, since it is important to establish that internal jet kinematics are consistently observed across all jet emission lines. 

\section{Observations}
\label{observations}

Spectroscopic observations were conducted using the Very Large Telescope (VLT) UT1 located at Paranal, Chile under observing program 077.C-0818A. The ISAAC instrument was employed to acquire data at near-infrared wavelengths from the bipolar outflow from the T\,Tauri star, Th\,28. The VLT/ISAAC spectrograph in the MR configuration with a 120$\times$0.3 arcsec$^2$ slit has a spatial sampling of 0$\farcs$147 and a spectral resolution of 8\,900-10\,500 at the investigated wavelength region. The latter translates to a velocity resolution of 29-33\,km\,s$^{-1}$. Where a line profile is intrinsically Gaussian, fitting allows an effective resolution which improves with climbing signal-to-noise, typically reaching one tenth of the resolution. Meanwhile, measurement of OH lines yield a deviation in absolute velocity calibration of $\sim$1.5\,km\,s$^{-1}$ with respect to the theoretical value. 

Data were obtained in service mode over three nights: 2006 April 14th, 19th and 20th. A log of the observations is given in Table\,\ref{obs_table}. Observations were conducted with a single slit placed parallel to the bipolar jet using three filters, with a seeing of $\sim$0$\farcs$5 (FWHM) during the observations. The {\it K} filter was centered on 2.12\,$\mu$m in order to capture H$_2$ emission at 2.12\,$\mu$m (i.e., at vacuum rest wavelength of 21\,218.36\,\AA). The {\it H} filter was centered on 1.62\,$\mu$m in order to capture [\ion{Fe}{2}] emission at both 1.60\,$\mu$m (i.e., at vacuum rest wavelength of 15\,999.086\,\AA) and 1.64\,$\mu$m (i.e., at vacuum rest wavelength of 16\,439.98\,\AA. The {\it J} filter was centered on 1.29\,$\mu$m in order to capture Pashen Beta emission (i.e., at vacuum rest wavelength of 12\,821.610\,\AA). Observations were also conducted with a single slit placed perpendicular to each lobe of the bipolar jet  at 0$\farcs$5 (i.e., 86\,AU deprojected distance) from the star itself. An offset of 0$\farcs$5 was requested, based on the instrument PSF and assuming a seeing of 0$\farcs$6. This arrangement was chosen to provide a clear spectrum transverse to the jet axis and close to the star but without contamination from star light. However, the actual seeing was larger. The receding lobe was observed in the {\it H} and {\it K} filters, with a seeing on the night of 0$\farcs$7 and 0$\farcs$8 respectively, while the approaching jet was observed in the {\it H} filter with seeing of 0$\farcs$7. We discuss the consequences of this in Section\,\ref{results}. The standard AB\,B'A' nodding technique was employed during observations to allow good sky subtraction while maximising available observing time. Using the optimum exposure time of 300\,s, a total of eight individual exposures were made in each filter, and co-added to achieve the required signal-to-noise.  

Data reduction was carried out using IRAF tools, and included cosmic ray rejection, flatfield correction using the lamp exposures, wavelength calibration using the telluric OH lines, distortion rectification using a star trace from the ESO archive, double sky subtraction (or AB subtraction) using the nodded exposures, telluric correction and flux calibration using the spectroscopic standard. A heliocentric radial velocity correction of +20\,km\,s$^{-1}$ for all observations was determined using IRAF tools. This was applied to all subsequent radial velocity measurements along with a correction of -5\,km\,s$^{-1}$ to negate the star's heliocentric velocity \cite{Graham88}.

Finally, we also present previously unpublished {\it HST}/STIS H$\alpha$ data along side the other \ion{H}{1} lines. These observations were conducted on 2002 June 22 (proposal ID 9435), with the slit placed perpendicular to each jet lobe at 0$\farcs$3 (52\,AU deprojected distance) from the star. An aperture of 52$\times$0.1 arcsec$^2$ was used with the G750M grating which gave a radial velocity sampling of 25\,km\,s$^{-1}$ and diffraction limited spatial resolution of 0$\farcs$1. Spectra were obtained of the blue- and red-shifted lobes, using exposure times of 2200 and 2700 s, respectively. They included several forbidden lines  (which were previously published; \cite{Coffey04}) along with H$\alpha$ which we present here. Data were calibrated through the standard {\it HST} pipeline, subtraction of the  stellar continuum was performed, and hot/dark pixels were removed. 
\begin{table*}
\begin{center}
\scriptsize{\begin{tabular}{lcccccc}
\tableline\tableline
Slit position				&Filter	&Central wavelength	&Spectral Domain		&Resolving Power	&Seeing	&Exposure time	\\
						&		&($\mu$m)			&($\mu$m)			&				&(arcsec)	&(s)				\\  
\tableline
Parallel to bipolar jet			&K 		&2.12				&2.06 - 2.18			&8 900			&0.5		&2400		\\
						&H 		&1.62				&1.58 - 1.66			&10 000			&0.6		&2400		\\
						&J 		&1.29				&1.26 - 1.32			&10 500			&0.5		&2400		\\
Perpendicular to red lobe		&K 		&2.12				&2.06 - 2.18			&8 900			&0.8		&2400		\\
						&H 		&1.62				&1.58 - 1.66			&10 000			&0.7		&2400		\\
Perpendicular to blue lobe	&H 		&1.62				&1.58 - 1.66			&10 000			&0.7		&2400		\\
\tableline
\tableline 
\end{tabular}}
\end{center}
\caption{Details of observations of the jet from T Tauri star ThA 15-28 (a.k.a. Th 28) conducted with VLT/ISAAC in service mode over 3 nights in April 2006.
\label{obs_table}}
\end{table*}

\section{Results}
\label{results}

\begin{figure*}
\begin{center}
\epsscale{3.0}
\plotone{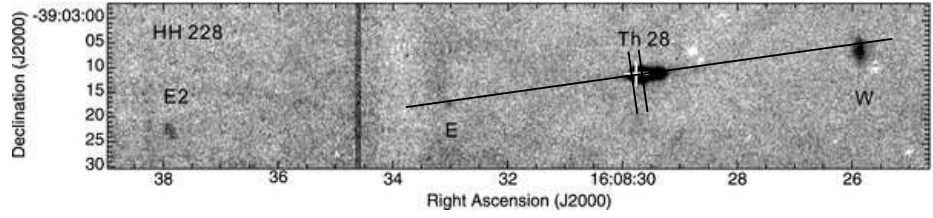}
\caption{Continuum subtracted [\ion{S}{2}] image of HH 228, adapted from \cite{Wang09}, with our VLT/ISAAC slit positions overlaid. 
\label{wang}}
\end{center}
\end{figure*}

The Herbig-Haro outflow from Th\,28, named HH 228, consists of a bipolar microjet close to the star, two bowshocks 39$\arcsec$ and 45$\arcsec$  either side of the star aligned close to east-west, and a third fainter bowshock farther to the east at 95$\arcsec$ from the star. The overall position angle of the flow is almost east-west, with the brighter red lobe of the microjet traveling towards the west, as illustrated in Figure\,\ref{wang}. The jet lies close to the plane of the sky, and consequently has low radial velocities of a few tens of km\,s$^{-1}$. High proper motions of 250 - 350 km\,s$^{-1}$ confirm that these objects are driven by Th 28 \cite{Wang09}. 

For the first time, we have detected the bipolar microjet in near-IR emission via spectra taken with the slit both parallel and perpendicular to the flow direction. We will first describe the 'parallel' configuration spectra, which illustrate the global morphological and kinematic properties of the flow close to the source, and then we will focus on the 'perpendicular' spectra, which highlight further important peculiarities. 

\subsection{Parallel-slit Configuration}

\begin{figure*}
\begin{center}
\epsscale{0.4}
\plotone{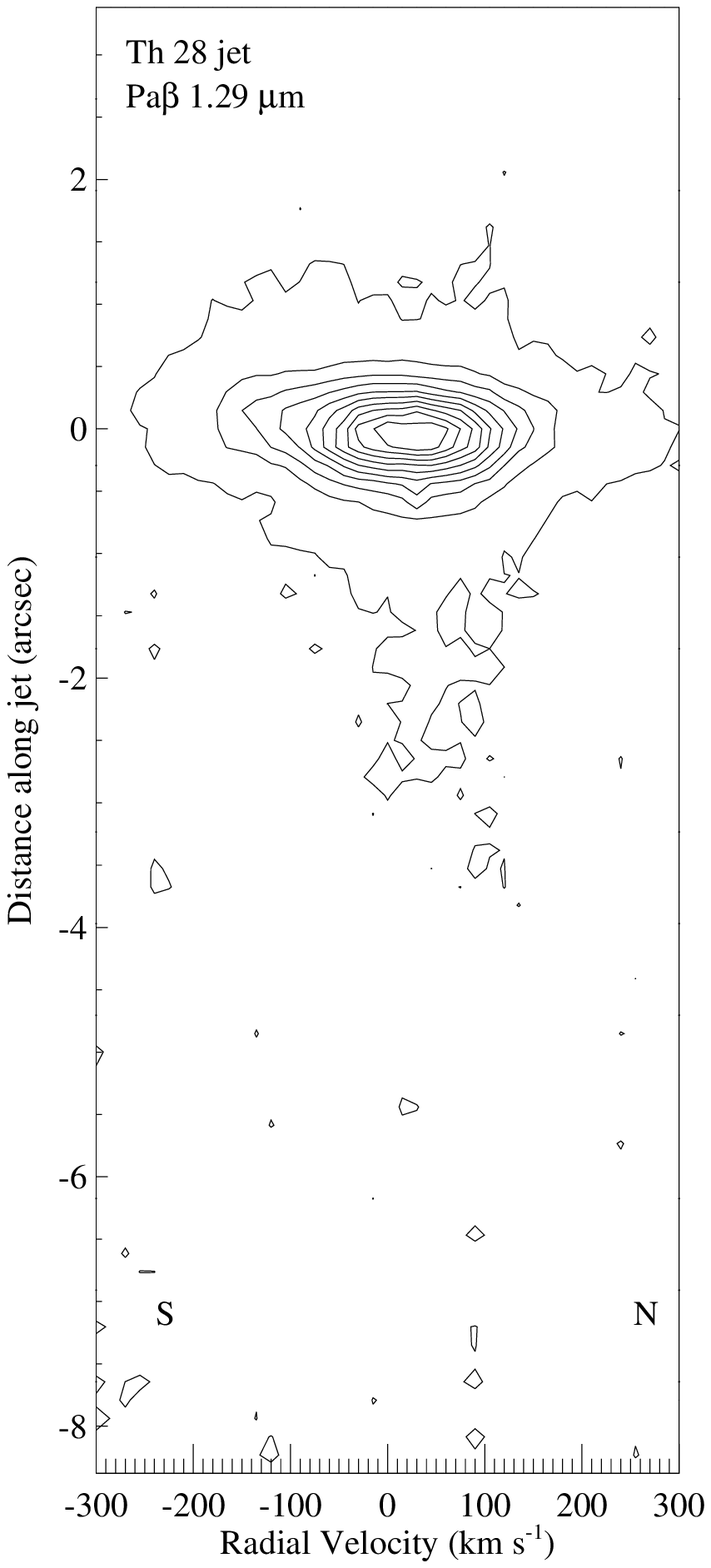}
\epsscale{0.9}
\plottwo{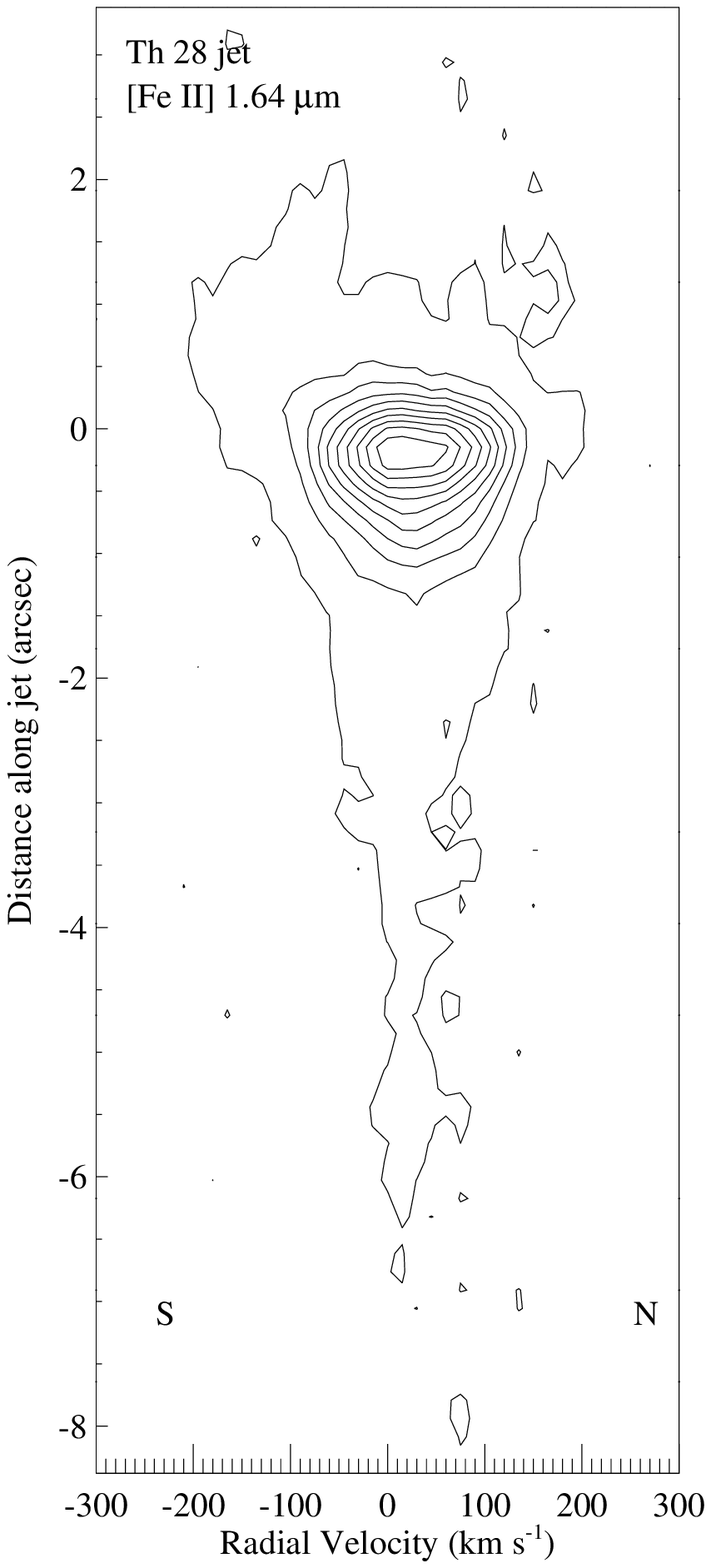}{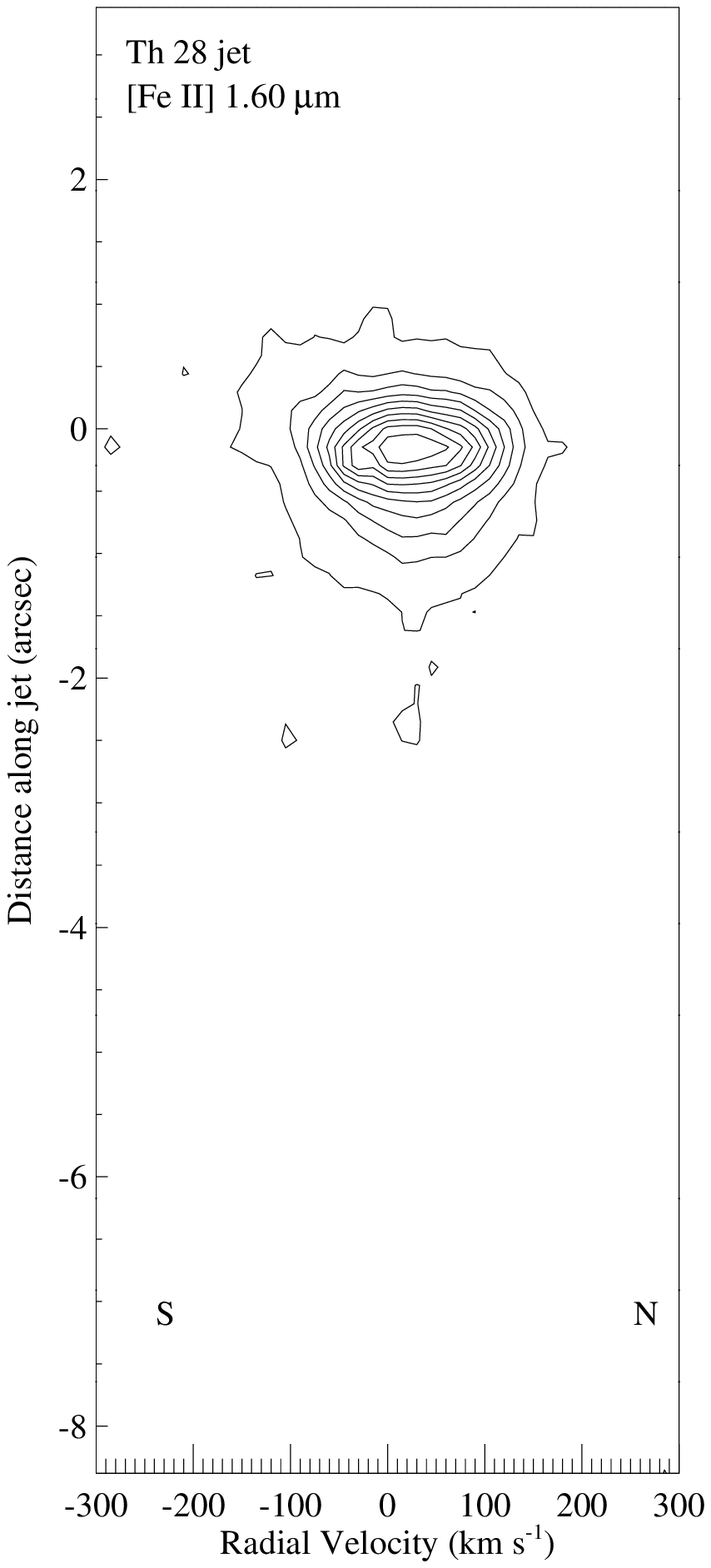}
\plottwo{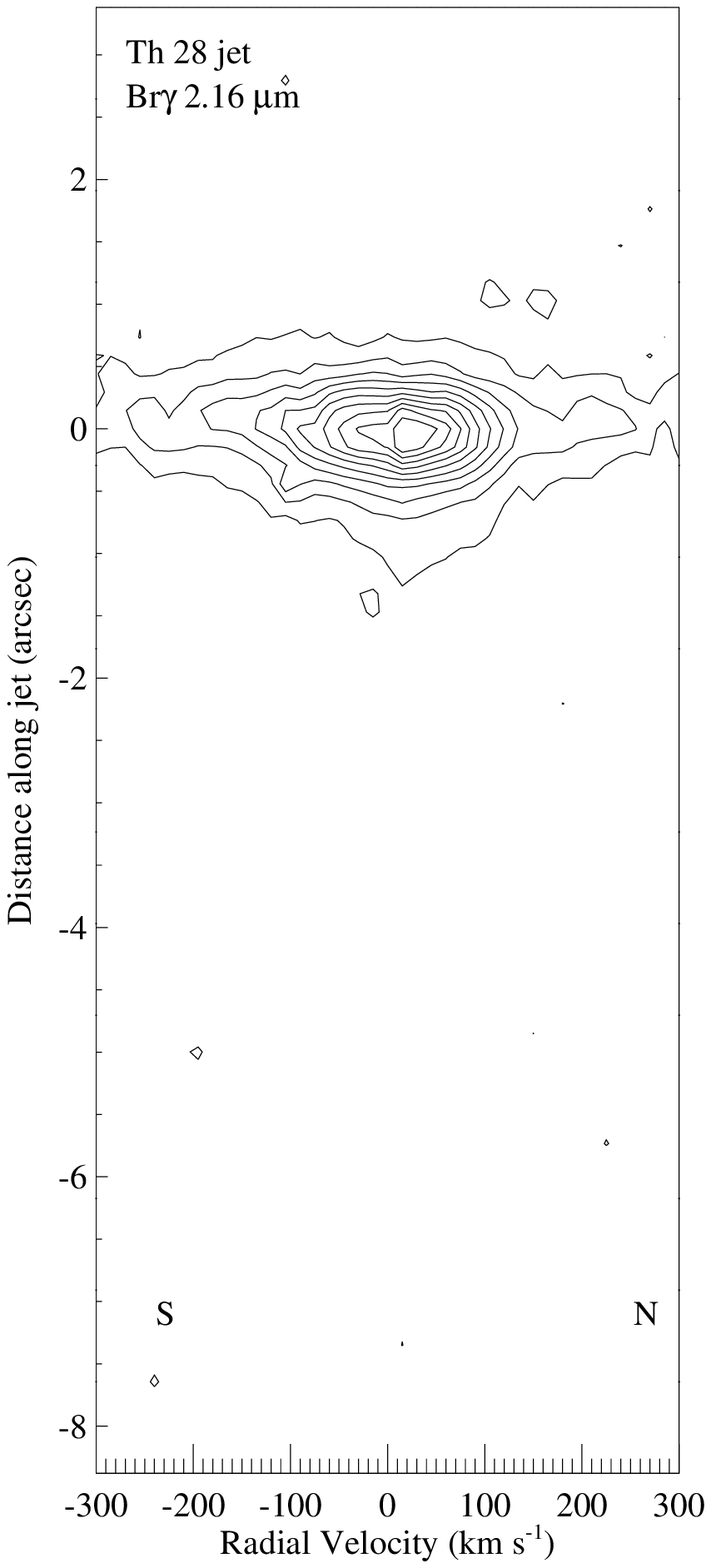}{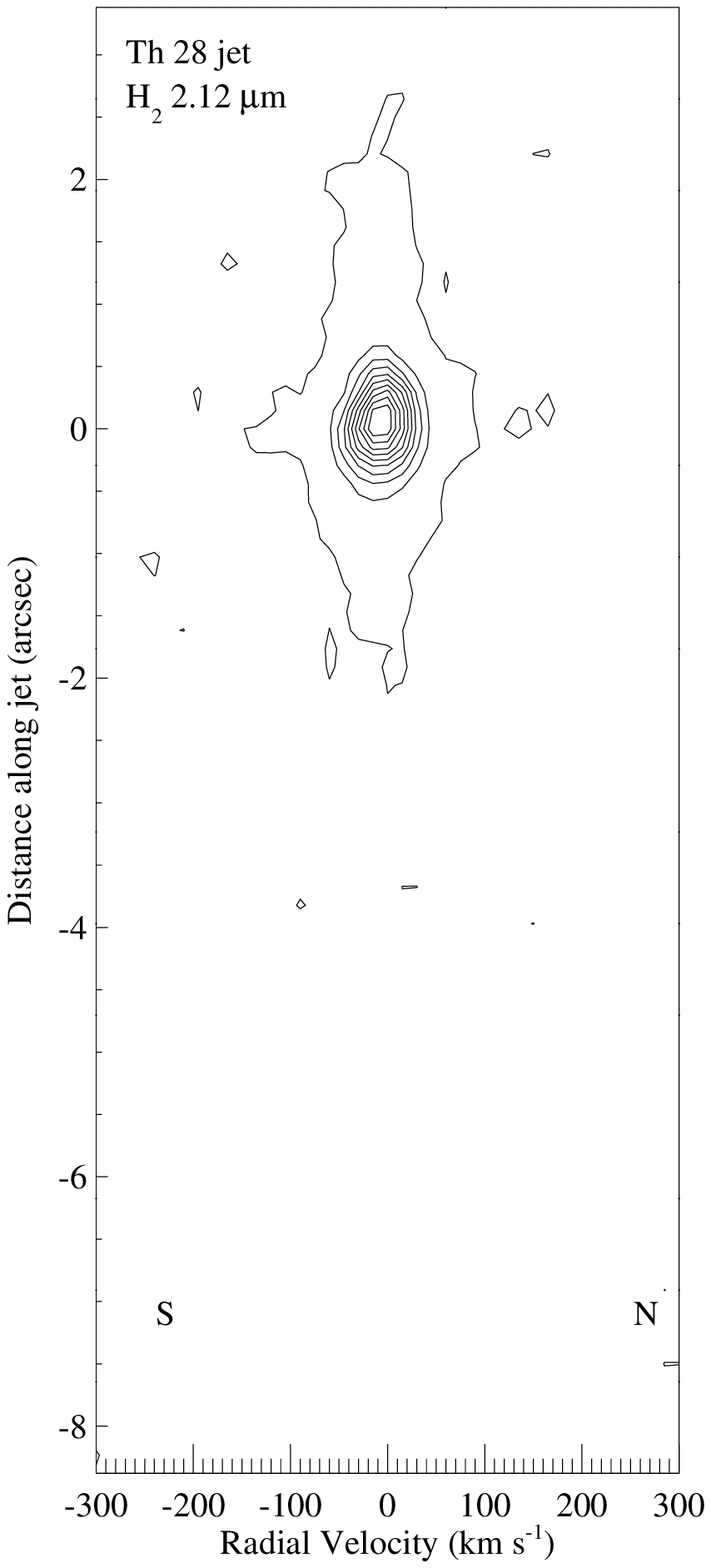}
\caption{PV plots of the Th 28 bipolar jet with the long slit parallel to the jet axis, illustrating the extent of emission with distance from the star. Each plot has ten linearly-spaced contour, with contour levels from 3$\sigma$ to 70, 890, 240, 70, 670$\sigma$ respectively. 
\label{para_pvs}}
\end{center}
\end{figure*}

In the parallel case, the position-velocity (PV) plots, collected in Figure\,\ref{para_pvs}, show detections in [\ion{Fe}{2}], Br$\gamma$, Pa$\beta$ and H$_2$. In these plots, the position of the central star was determined from a Gaussian fit to the continuum, which was then subtracted. 

The various emission lines all show a wide emission peak around the star, but since they probe different physical conditions, they extend to different distances from the source. In the atomic lines, the red lobe is brighter, and corresponds to negative distances along the jet in Figure\,\ref{para_pvs}. [\ion{Fe}{2}] traces both lobes of the bipolar jet, with [\ion{Fe}{2}]1.64\,$\mu$m emission from the red lobe being stronger and persisting along the jet to 6$\arcsec$, compared to 2$\arcsec$ for the blue lobe. The Pa$\beta$ emission clearly extends to 3$\arcsec$ while Br$\gamma$, which is intrinsically fainter than Pa$\beta$, is closer to the jet base and we do not see it beyond 1.5$\arcsec$ along the receding flow. The fact that Pa$\beta$ and Br$\gamma$ are clearly seen at a distance from the source tracing the red lobe in the PV diagrams is very interesting, as these lines have been often commonly attributed to accretion phenomena. Meanwhile, the H$_2$ emission presents a different behaviour with respect to the atomic lines: it is more confined to a narrow range of low velocities, and is more extended on the side of the blue lobe.

\begin{figure*}
\epsscale{2.0}
\plottwo{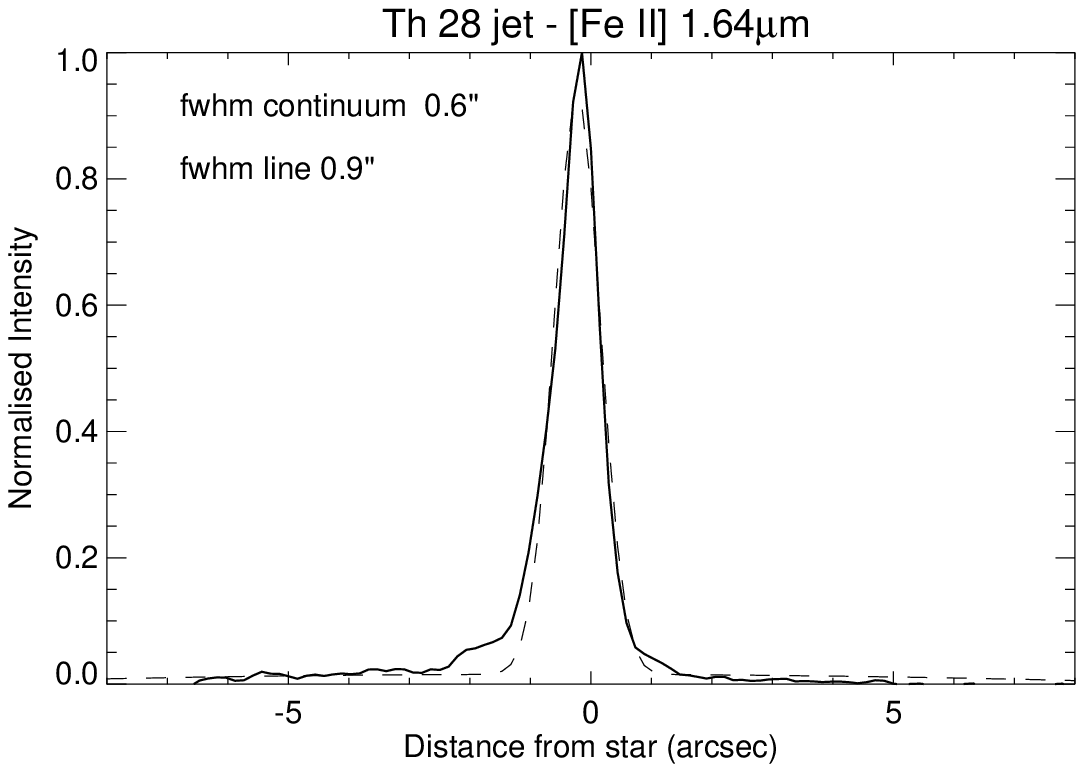}{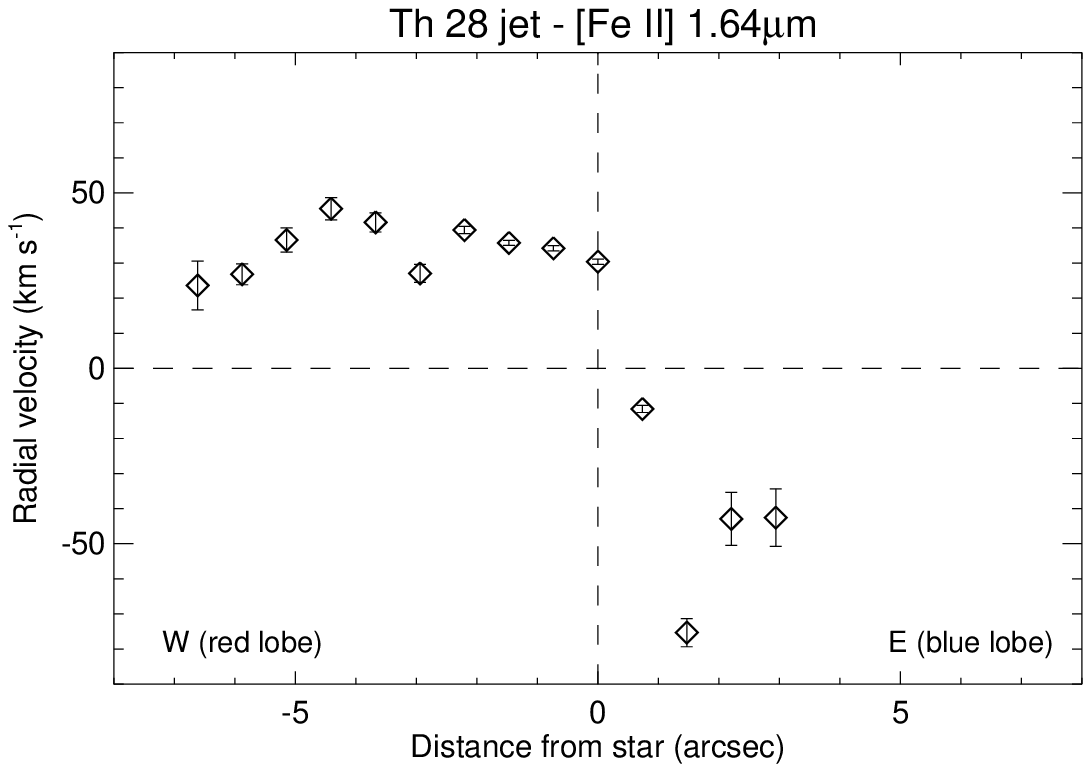}\\
\plottwo{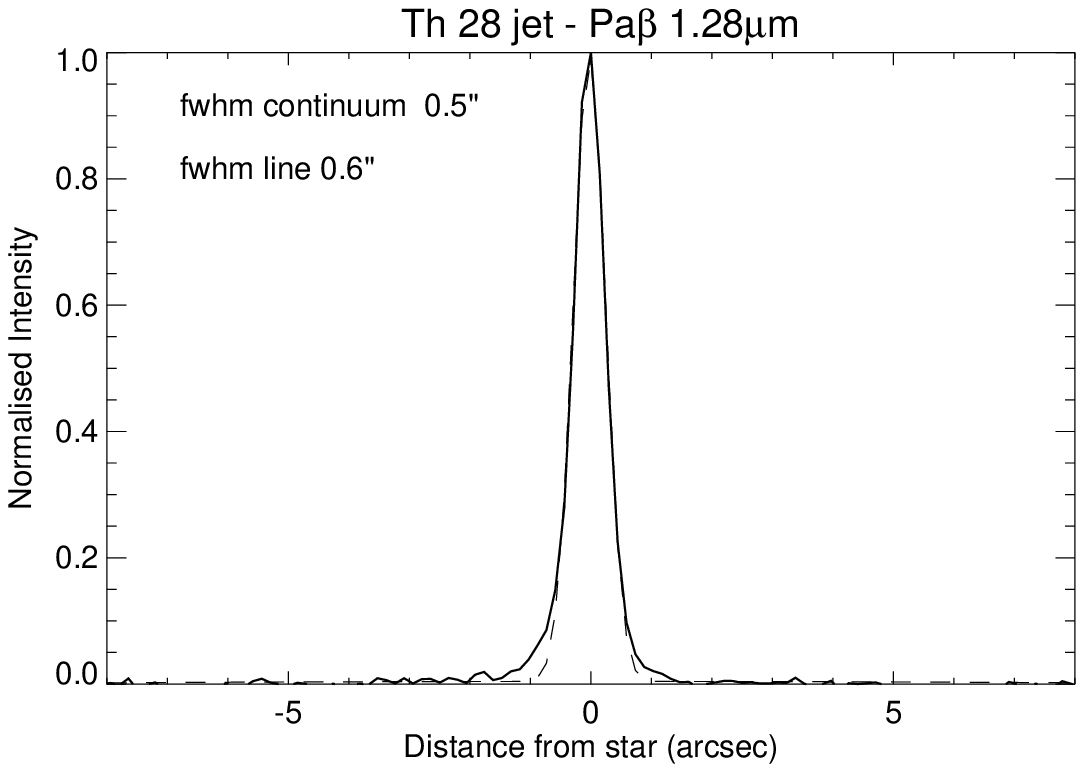}{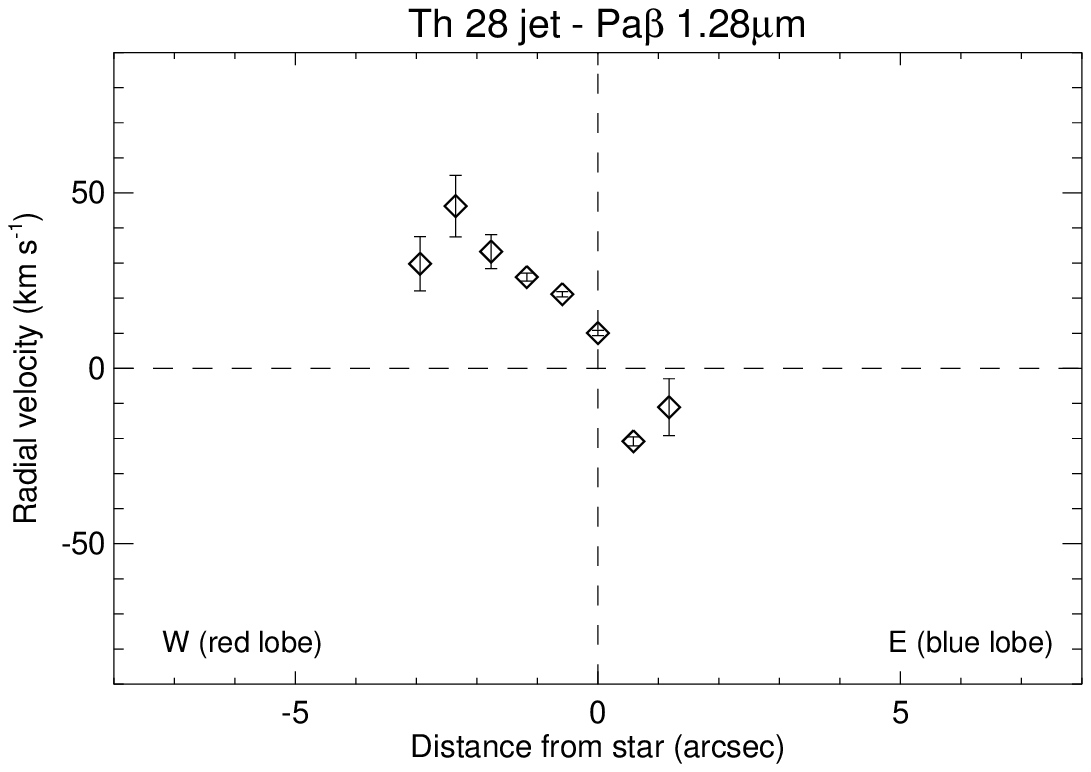}\\
\plottwo{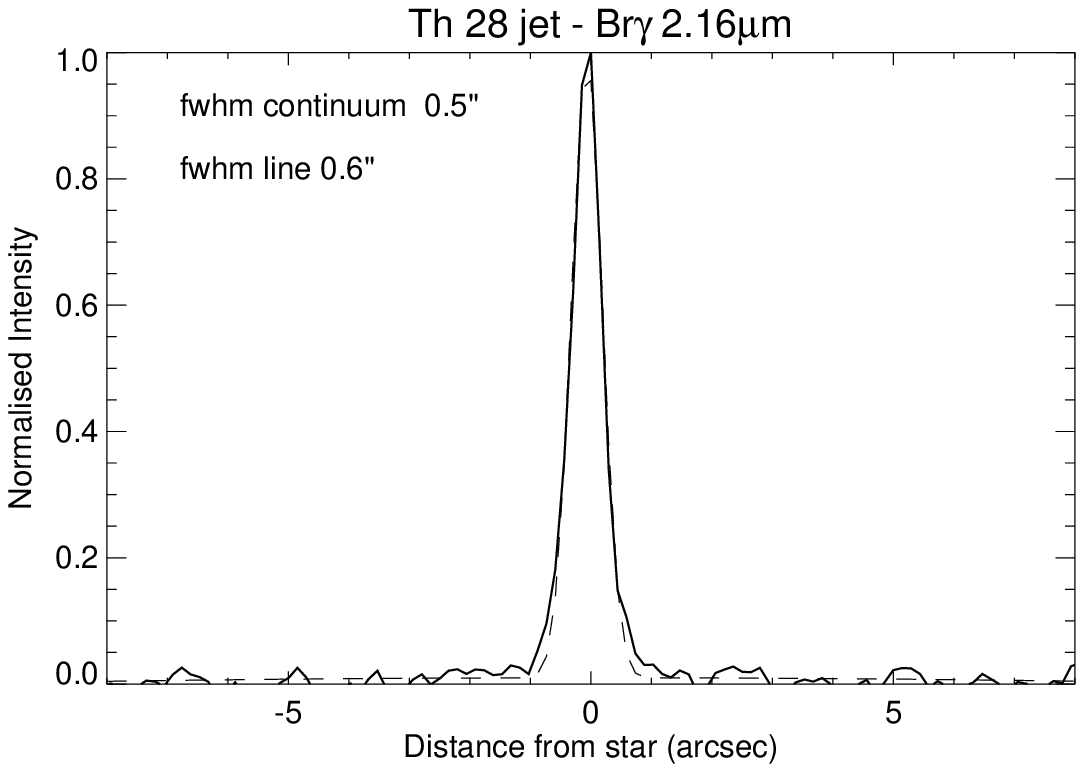}{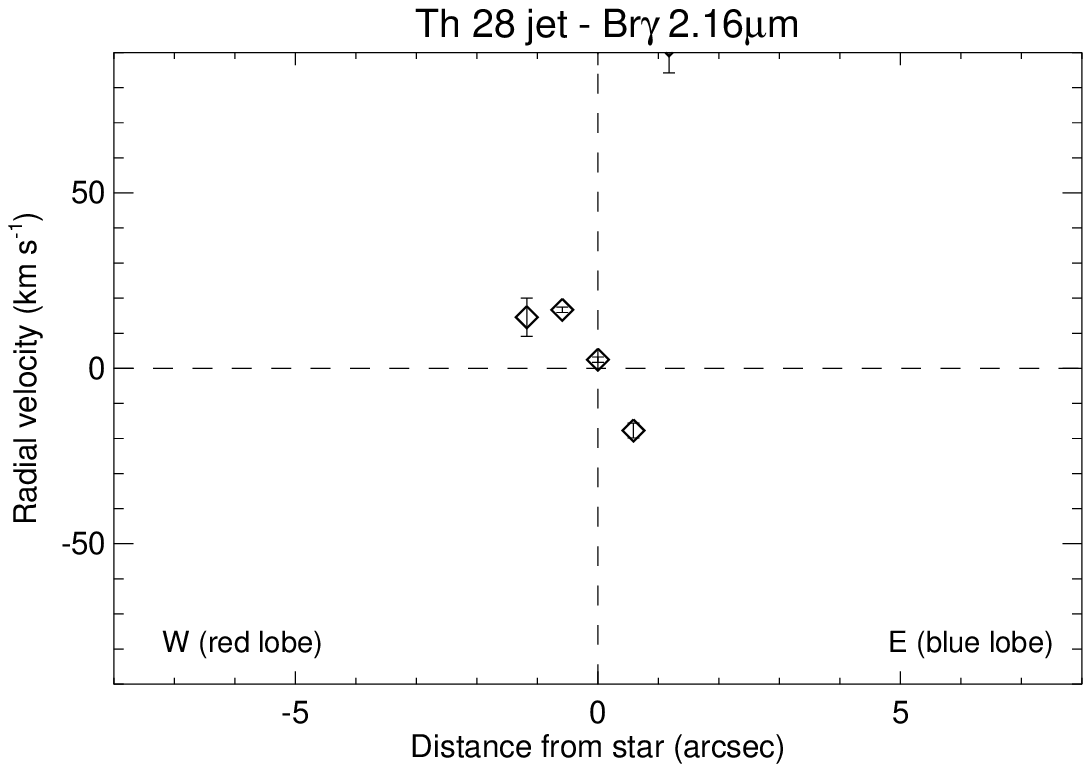}\\
\plottwo{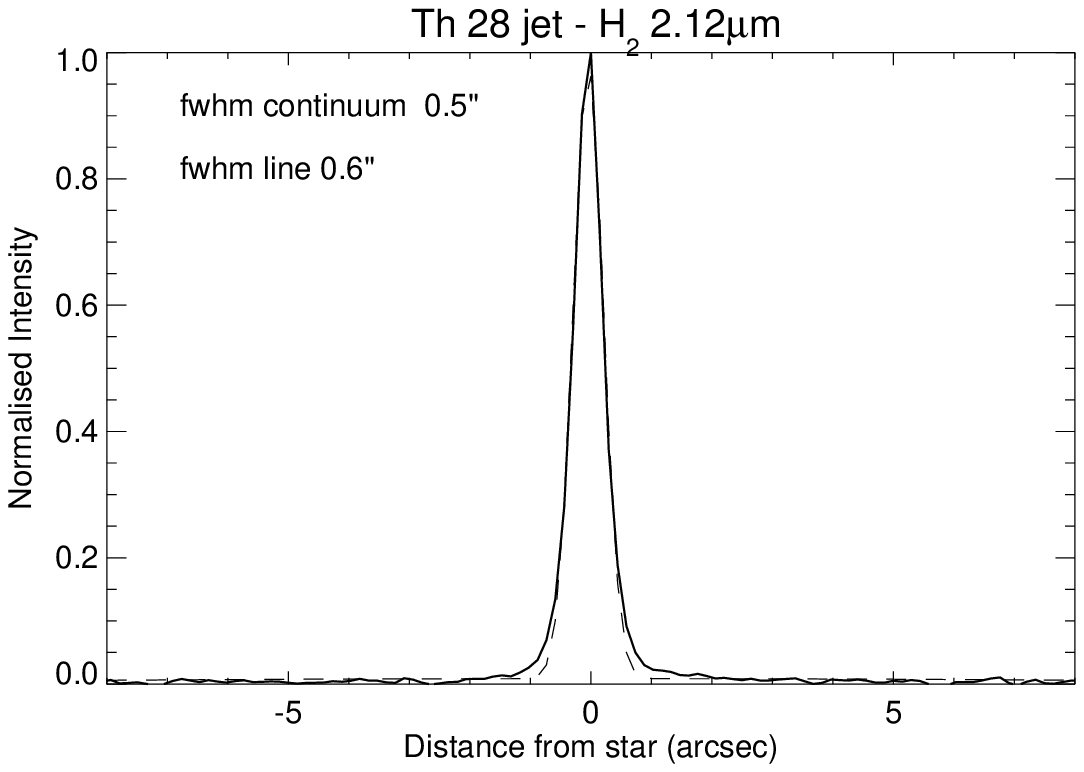}{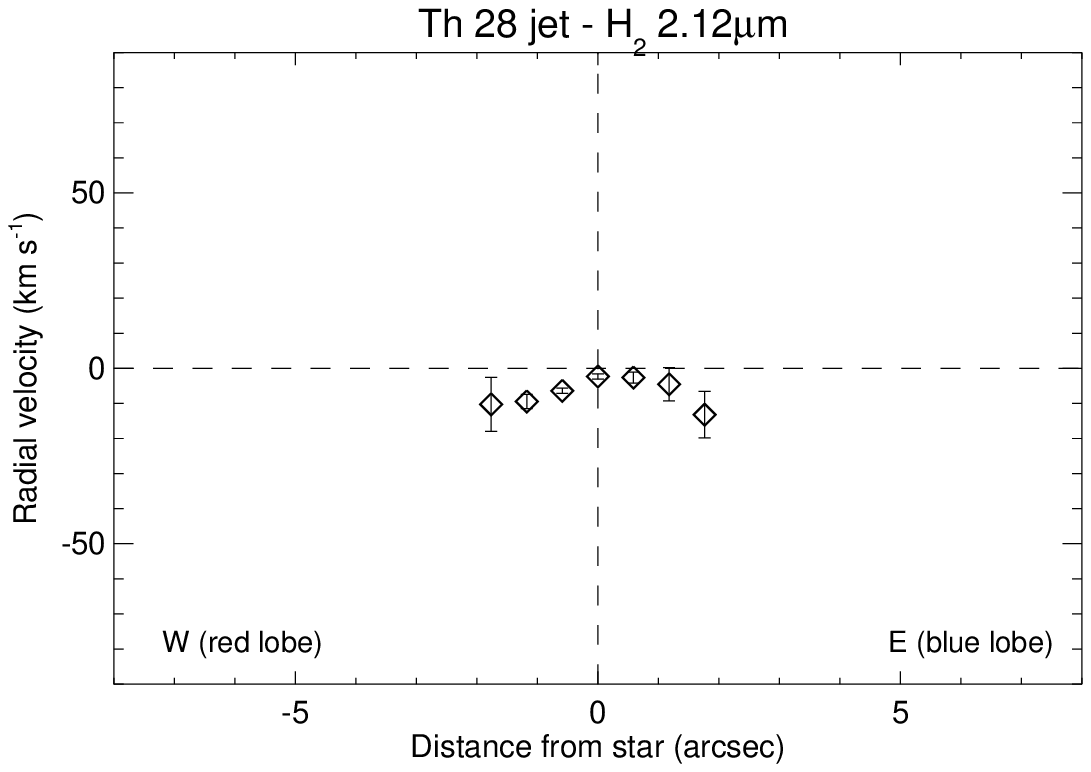}\\
\caption{Spatial intensity and velocity profiles of Th 28 bipolar jet in near-IR lines through a slit parallel to the flow axis. The wings of the profile extend outside a simple Gaussian fit to the emission (dashed line) indicating an extended source. Flux is spatially binned to match the seeing. 
\label{para_profiles}}
\end{figure*}

To further examine the structure and kinematics indicated by these lines, the flux was spatially binned to match the seeing. 
The spatial profile wings extend beyond a simple spatial Gaussian fit of the line emission (dashed curve) 
implying origin in an extended source, Figure\,\ref{para_profiles} left column. Corresponding velocity profiles along the jet were determined from Gaussian fits in the velocity direction performed at different distances from the source, Figure\,\ref{para_profiles}, right column. We see that [\ion{Fe}{2}] travels with a velocity of about +30\,km\,s$^{-1}$ in the red lobe and -50\,km\,s$^{-1}$ in the blue lobe. The plots also show that Br$\gamma$ and Pa$\beta$ travel with similar velocity to the well known atomic jet tracer [\ion{Fe}{2}]. These permitted H lines appear to be produced in the jet to distances of $\pm$0$\farcs$6 from the star for Br$\gamma$, and from -3$\arcsec$ to +1$\farcs$2 from the star for Pa$\beta$. In the same CCD frame with Pa$\beta$, four other iron lines of similar excitation to [\ion{Fe}{2}] 1.64 $\micron$, but lesser intensity, were also observed in the parallel slit, namely [\ion{Fe}{2}] 12706.94, 12791.255, 12946.232 and 12981.279 \AA (vacuum wavelengths). All of these show velocity profiles consistent with that of the [\ion{Fe}{2}] 1.64 $\micron$ line. More significantly, the higher excitation [\ion{Fe}{2}] 2.13$\micron$ line was also detected. This is an important confirmation that the similarly high excitation hydrogen lines, Br$\gamma$ and Pa$\beta$, originate in the jet itself. 

Finally, we find that the H$_2$ emission unexpectedly shows blue-shifted velocities of about -10\,km\,s$^{-1}$ in the direction of both the receding and approaching jet lobes. Its spatial distribution is also different to the atomic lines, in that it is more evenly distributed about the system, and slightly more extended toward the blue lobe. In addition, the H$_2$ line is narrower than the atomic lines, indicating that the emitting material here is colder and/or less affected by turbulent motions. 

\subsection{Perpendicular-slit Configuration}
\label{Perslitconfig}

\begin{figure*}
\begin{center}
\epsscale{2.0}
\plottwo{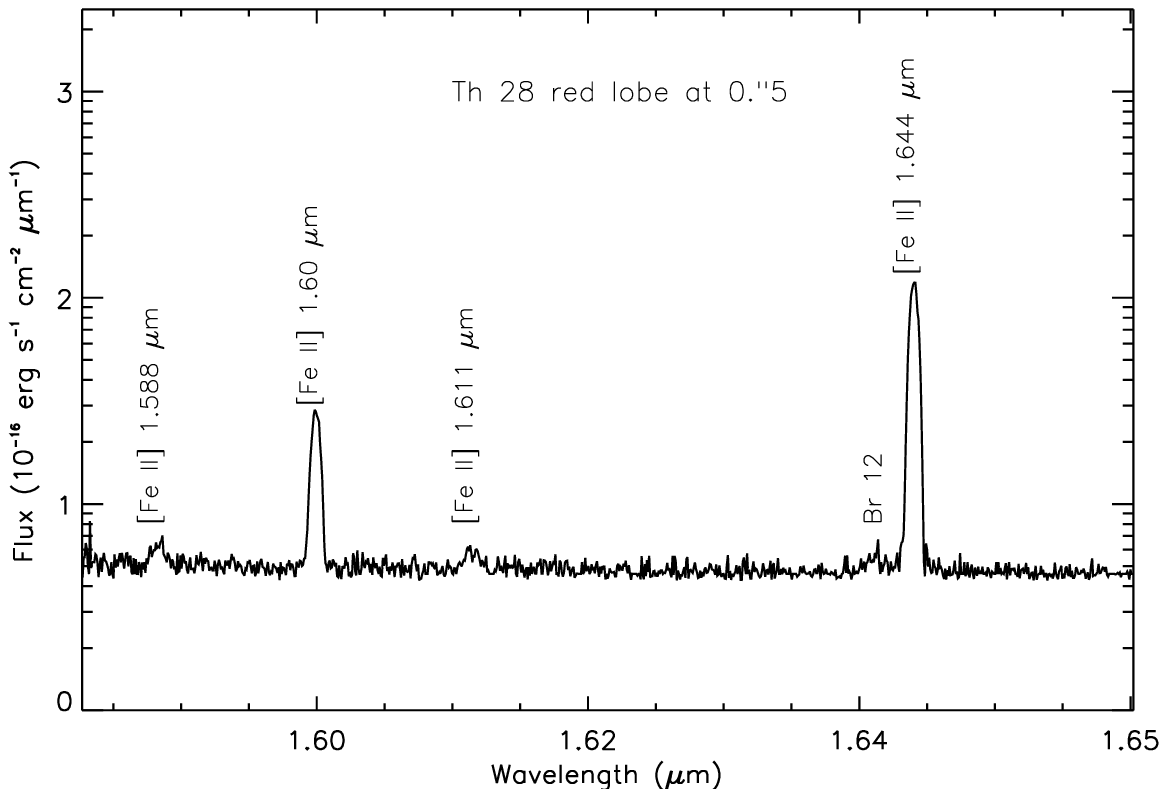}{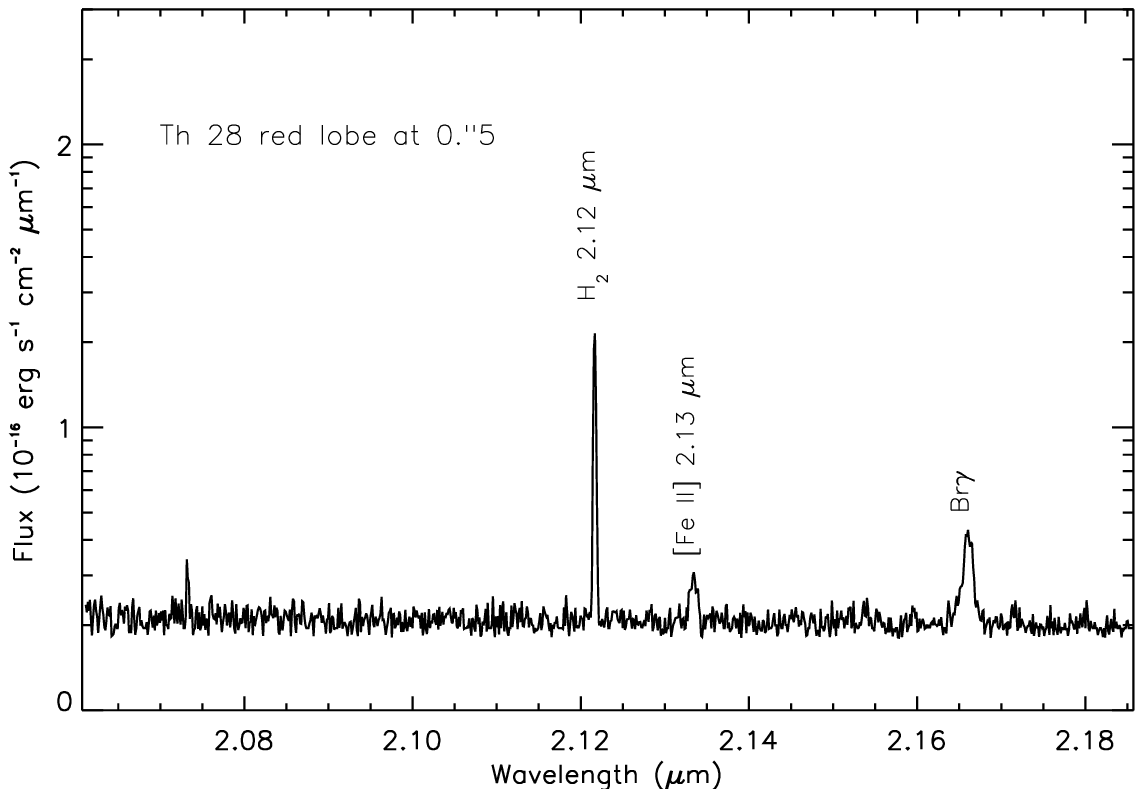}
\epsscale{1.0}
\plotone{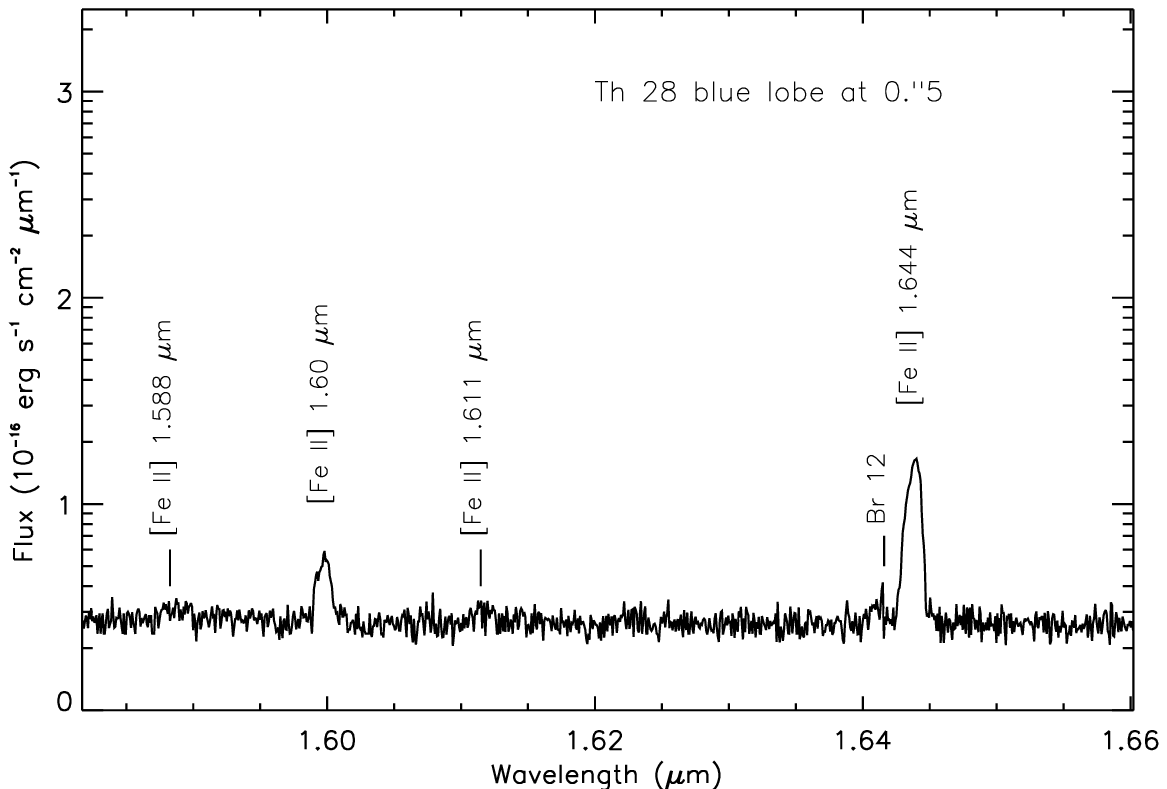}
\caption{Extracted one-dimensional spectra of near-IR atomic and molecular lines, derived from jet spectra with the slit perpendicular to jet propogation. The spectra reveal single-peaked emission in H$_2$, [\ion{Fe}{2}] and Br$_{\gamma}$ originating in the region of the microjet. The H$_2$ emission is stronger but much narrower, implying origin in a medium colder than the atomic lines. 
\label{onedspec}}
\end{center}
\end{figure*}

To understand in further detail the conditions close to the jet base, we examine each jet lobe via our data taken with the slit perpendicular to the jet axis at a distance of 0$\farcs$5 from the source. In Figure\,\ref{onedspec} we first show 
flux-calibrated one-dimensional extracted spectra, that globally present the detections in near-IR molecular and atomic lines. We  find  H$_2$, [\ion{Fe}{2}] lines and again, interestingly, Br$_{\gamma}$. 
While both lobes were detected in [\ion{Fe}{2}] lines, emission was fainter for the approaching jet than for the receding jet, highlighting the fact that the former is more embedded. Note that the spectral FWHM of the H$_2$ line in the receding lobe is much smaller than that of the other lines, in agreement with the findings for the 'parallel' spectra. Perpendicular slit H$_2$ observations were not conducted for the approaching jet.  

The main emission properties derivable form extracted spectra are collected in 
Table\,\ref{onedspec_table}, where for each line we list the calibrated flux and
signal-to-noise, the velocity of the emission peak and the velocity FWHM, and finally the jet spatial
width. The jet width was resolved only in H$_2$ (FWHM of 1$\farcs$6) 
under the seeing during the observations of not better than 0$\farcs$7. 
The fact that the atomic near-IR lines were spatially unresolved 
precluded any conclusions on jet rotation for these lines, since 
this would be determined from radial velocity gradients detected across the jet width. 
Nevertheless, the available near-IR perpendicular spectra were very useful in identifying relevant physical conditions in the jet.

\begin{table*}
\begin{center}
\scriptsize{\begin{tabular}{llcccccc}
\tableline\tableline
Target		&Emission			&Flux			&FWHM		&S/N		&$v$			&Seeing		&Jet width		\\
		&($\mu$m)	&(10$^{-16}$\,erg\,cm$^{-2}$\,s$^{-1}$)	&(km\,s$^{-1}$)	&		&(km\,s$^{-1}$)		&(arcsec)	&(arcsec)\\  
\tableline
Red lobe	&H$_2$\,$\lambda$2.12 		&7.1~$\pm$0.1		&45		&112		&-12			&0.8		&1.6		\\
		&[Fe II]\,$\lambda$2.13 	&1.0~$\pm$0.3		&138		&6		&+23			&0.8		&unresolved		\\
		&Br$\gamma$$\lambda$2.16	&3.1~$\pm$0.3		&187		&11		&+6			&0.8		&"		\\
		&[Fe II]\,$\lambda$1.64		&82.5~$\pm$0.3		&134		&315		&+24			&0.7		&"		\\
		&[Fe II]\,$\lambda$1.60		&16.7~$\pm$0.3		&139		&69		&+18			&0.7		&"		\\
Blue lobe	&[Fe II]\,$\lambda$1.64		&17.1~$\pm$0.3		&215		&49		&-16			&0.7		&"		\\
		&[Fe II]\,$\lambda$1.60		&3.1~$\pm$0.3		&215		&12		&-13			&0.7		&"		\\
\tableline
\tableline 
\end{tabular}}
\end{center}
\caption{Intensity and kinematical information derived from near-IR emission from the Th\,28 bipolar jet obtained with a slit perpendicular to the flow direction at 0$\farcs$5 from the star. The jet width is spatially unresolved in all atomic lines, contrary to the H$_2$ emission. Fluxes may be over-estimates, and velocities under-estimates due to scattered light (see Section\,\ref{Perslitconfig}). 
\label{onedspec_table}}
\end{table*}

\begin{figure*}
\begin{center}
\epsscale{1.5}
\plottwo{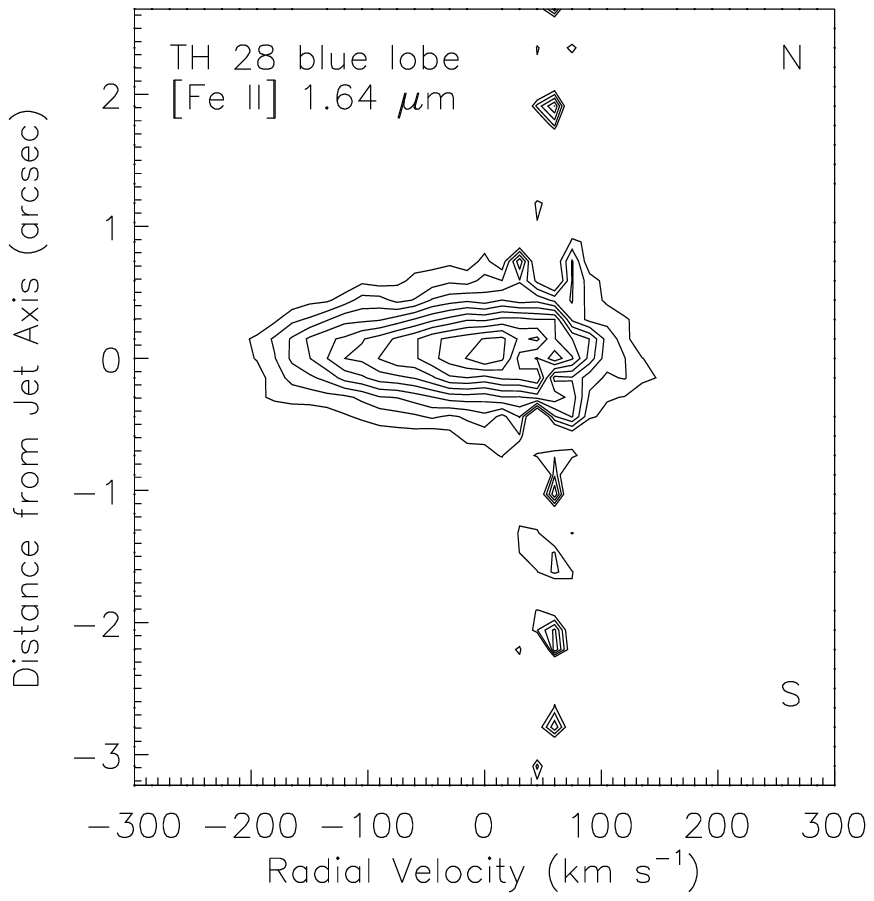}{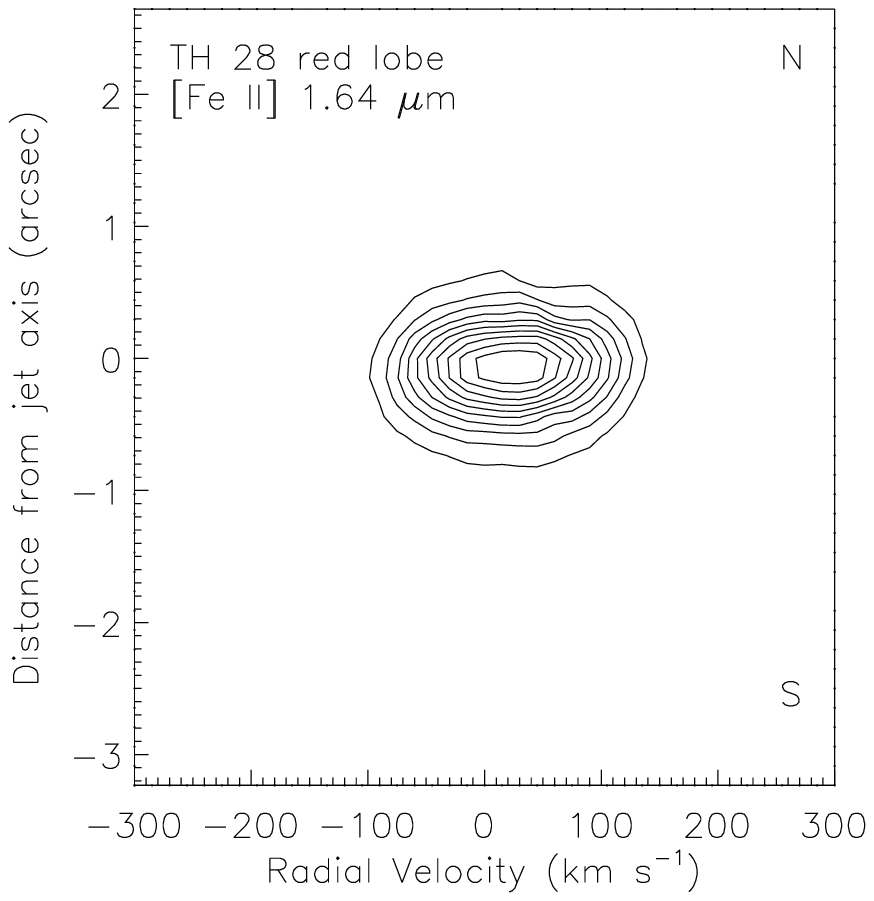}
\epsscale{0.65}
\plotone{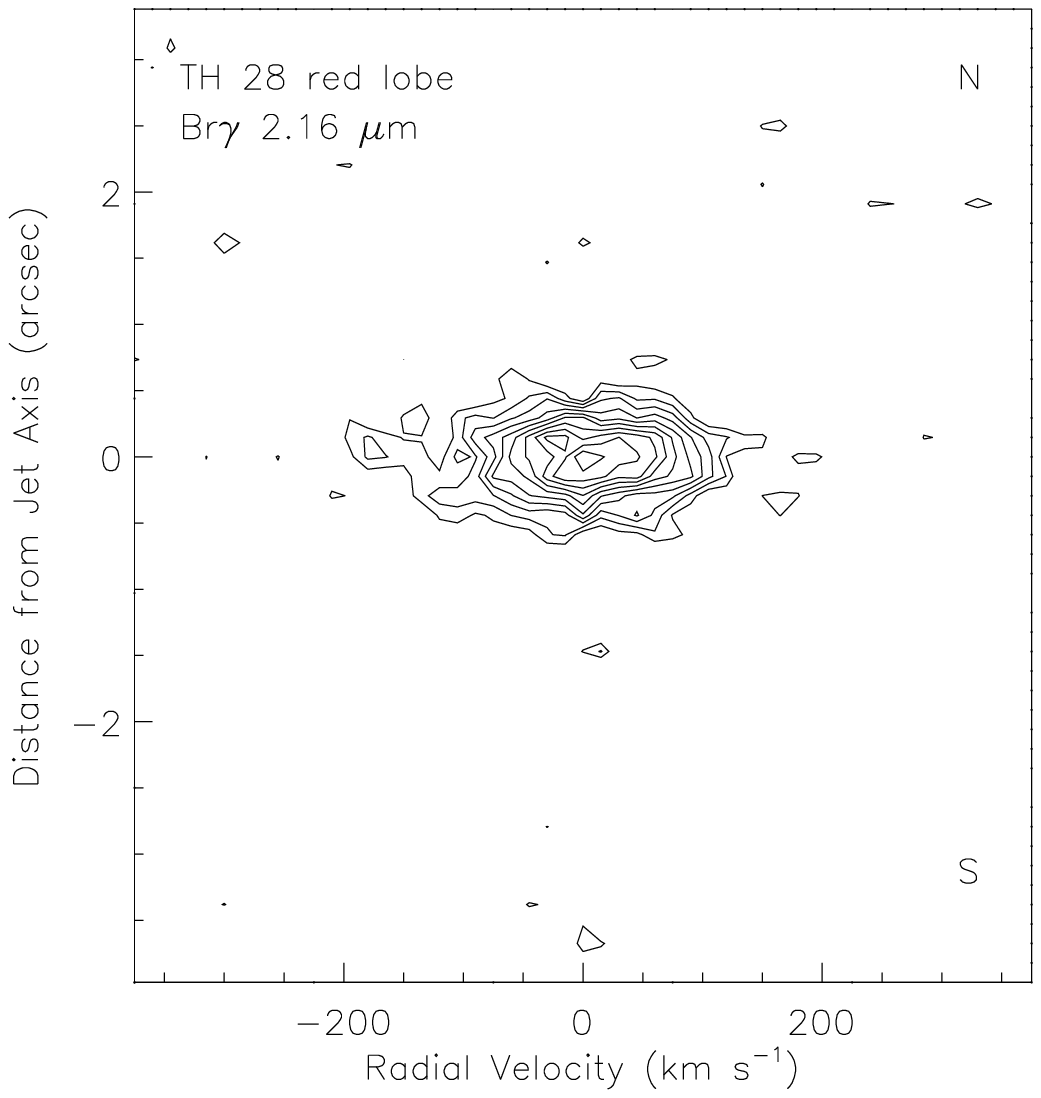}
\caption{PV plots of the Th 28 bipolar jet with the long slit perpendicular to the jet axis. Each plot has ten linearly-spaced contour, with contour levels from  3 to 60, 180 and 20$\sigma$ respectively. A night sky line disrupts the [\ion{Fe}{2}] emission. 
\label{perp_pv}}
\end{center}
\end{figure*}

\begin{figure*}
\begin{center}
\epsscale{1.3}
\plottwo{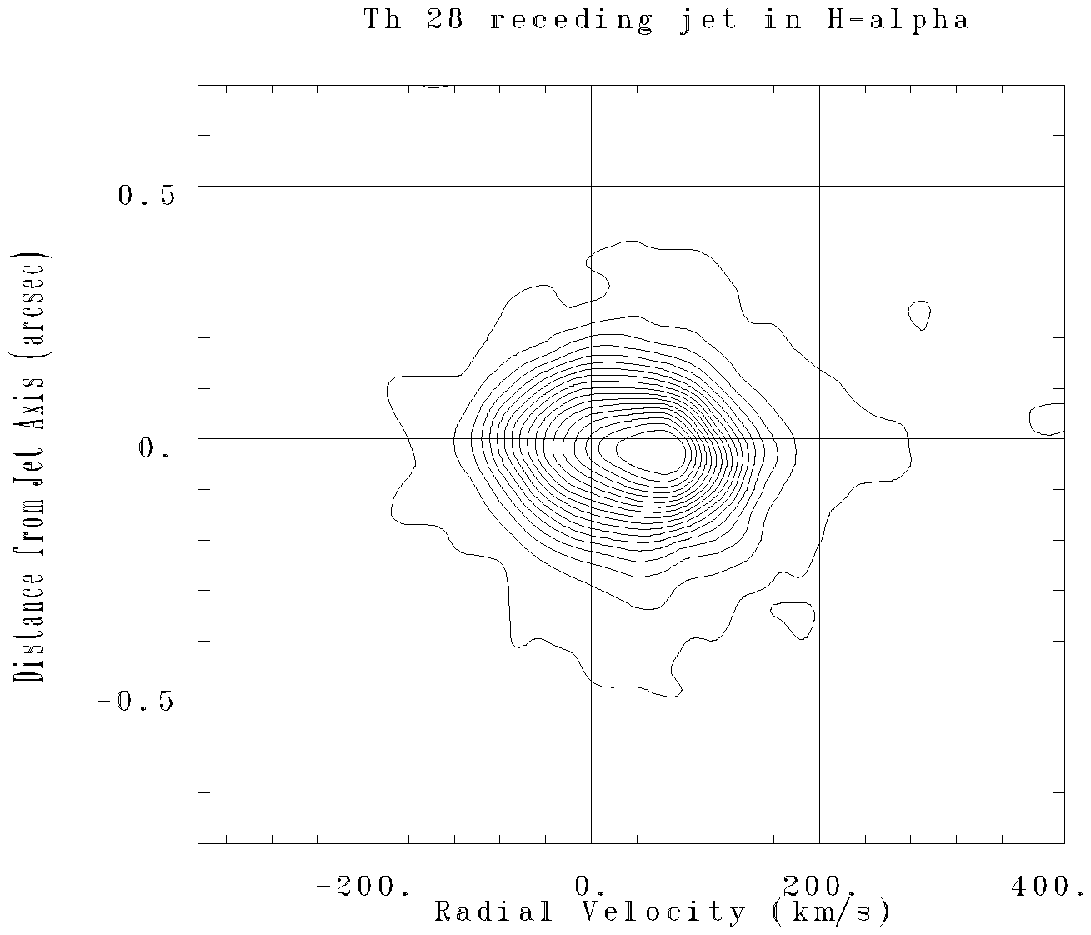}{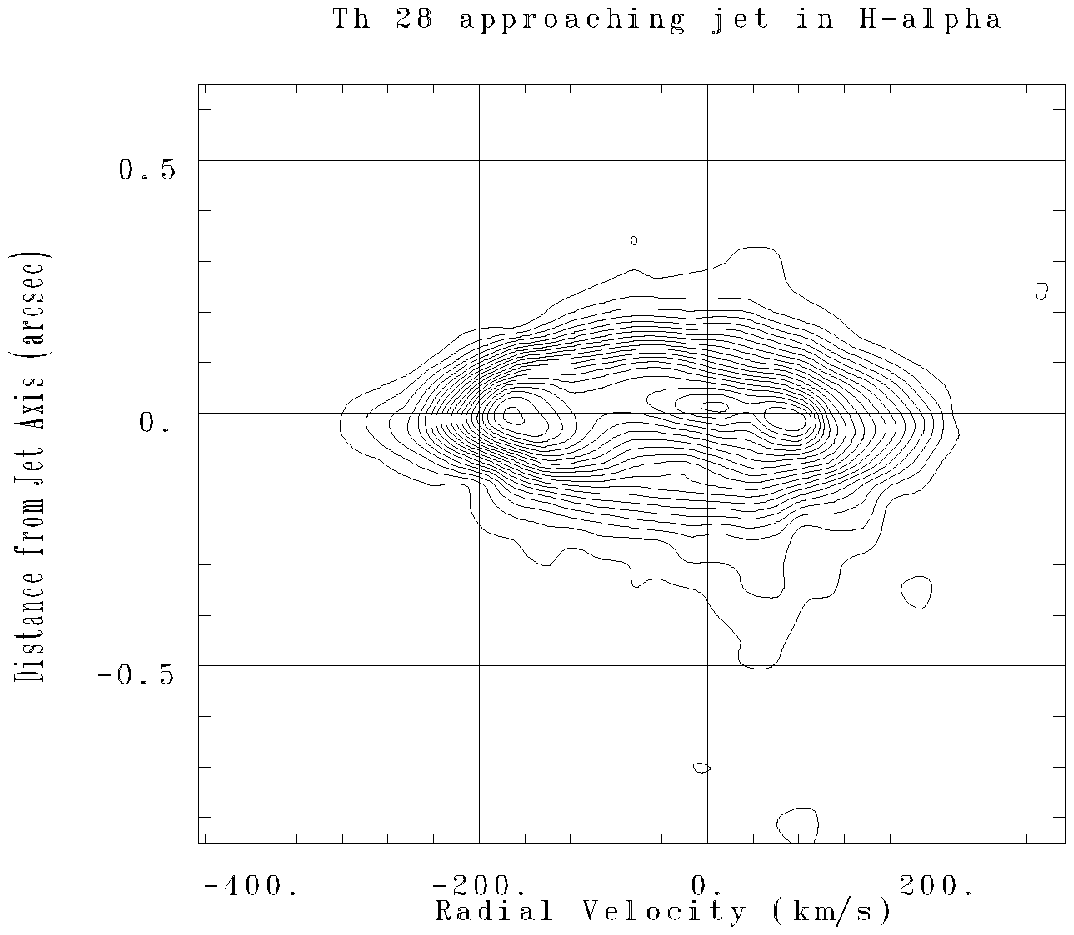}
\caption{Th 28 bipolar jet observed in H$\alpha$ via {\it HST}/STIS at 0$\farcs$3 from the star, with a long slit positioned perpendicular to the flow direction. Contours in units of erg\,cm$^{-2}$\,s$^{-1}$\,\AA$^{-1}$\,arcsec$^{-2}$  are from 3$\times$10$^{-15}$ to 1.22$\times$10$^{-13}$ and 6.0$\times$10$^{-14}$ respectively.
\label{perp_pv_ha}}
\end{center}
\end{figure*}

The information of Table\,\ref{onedspec_table} is further complemented by Figure\,\ref{perp_pv}, where we present the PV diagrams of [\ion{Fe}{2}] 1.64 $\mu$m line in both lobes and Br$_{\gamma}$ in the red lobe. For comparison with optical emission, in Figure\,\ref{perp_pv_ha} we also present a PV plot of the H$\alpha$ emission in {\it HST}/STIS spectra, taken with a similar instrument configuration, but as yet unpublished. 

The PV diagrams reveal a clear velocity asymmetry between the red and blue lobe. The high-velocity wing of the iron line in the blue lobe reaches -200\,km\,s$^{-2}$ while the red lobe reaches +120\,km\,s$^{-2}$. The asymmetry is reflected in Table\,\ref{onedspec_table} which reports the peak velocities via Gaussian fits to the integrated velocity profiles, and yields +25\,km\,s$^{-1}$ for the receding jet and -15\,km\,s$^{-1}$ for the approaching jet [\ion{Fe}{2}] emission. A similar behaviour is observed in the optical lines measured with {\it HST}/STIS, Figure\,\ref{perp_pv_ha}. In this case too, the emission on the blue side is fainter but it extends to higher velocities, with limiting values similar to the iron lines. The peak velocities and resulting asymmetry for H$\alpha$ are greater than the near-IR lines however, with values measured as -127\,km\,s$^{-1}$ and +62\,km\,s$^{-1}$ for the blue and red lobes respectively. 

On closer examination of the PV plots, it is clear that near-IR atomic emission covers a broad velocity range either side of zero. This is confirmed by a comparison of the peak velocity with the FWHM values of Table\,\ref{onedspec_table}, a spread which is too large to be explained by the velocity resolution. Indeed, the H$\alpha$ PV plot shows that in the blue lobe the emission appears to be double-peaked, with two distinct velocity components, one blueshifted and one redshifted. A double Gaussian fit to the profile reveals velocities consistent with the red and blue lobe velocities. It seems that the slit positioned over the blue lobe also takes in reflected light from the red lobe. This is further confirmed by the spatial asymmetry of the redshifted peak which matches exactly with the obvious spatial asymmetry of the red lobe PV plots in [\ion{O}{1}] \cite{Coffey08}, and is also coincident with an electron density asymmetry across the axis \cite{Coffey08}. A similar transverse spatial asymmetry could not be identified in the bright [\ion{Fe}{2}] 1.64 $\micron$ line, due to overlap with a night sky line. The scattered light is likely to be caused by the nebulosity around the blue lobe. Note that there is a less obvious reflection in the slit of the red lobe, which is compatible with the blue lobe being more embedded. This reflection from the opposing lobe was not encountered in the other targets examined under these {\it HST} proposals with a perpendicular slit configuration (proposal ID 9435; 9807). The implications for the parallel slit spectra are simply that any contamination from the opposing lobe will result in an under-estimation of the radial velocity values and an overestimation of the flux, but there are no implications for the interpretation. 
Indeed, the main concern here is the true origin of Br$_{\gamma}$ in the perpendicular-slit "jet" spectrum. Our slit was positioned at 0$\farcs$5 from the star in each lobe, to allow for the instrument PSF and to avoid stellar flux. Unfortunately, since the seeing on the night was in fact 0$\farcs$7 - 0$\farcs$8, the spectra unavoidably include contributions aside from the jet itself. As expected, stellar continuum emission is seen in these spectra, but was subtracted prior to analysis of the lines. Meanwhile, we overcome the difficulty of scattered light by relying on velocity measurements to indicate whether the emission traces the jet. The Br$_{\gamma}$ velocity of +6\,km\,s$^{-1}$ in the receding jet, with a velocity error of 1.5\,km\,s$^{-1}$, combined with a detection of [\ion{Fe}{2}] 2.13$\micron$ emission in the same region, strongly argues that the Br$_{\gamma}$  line truly trace outflowing gas. 

\begin{figure*}
\begin{center}
\epsscale{0.6}
\plotone{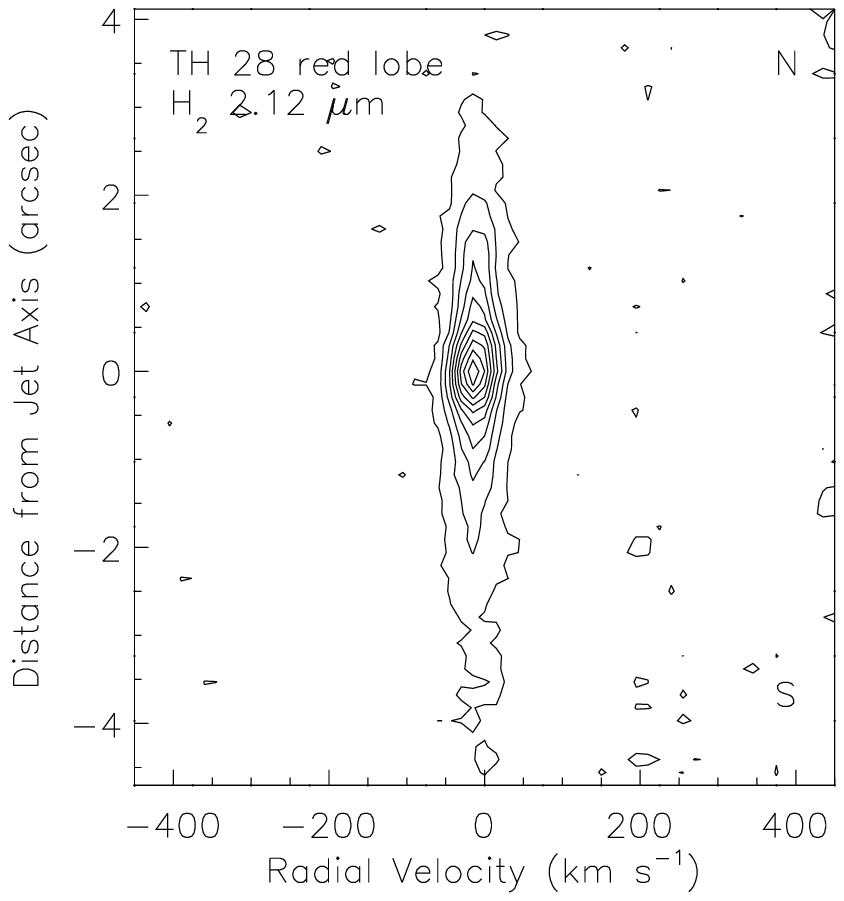}
\epsscale{1.65}
\plottwo{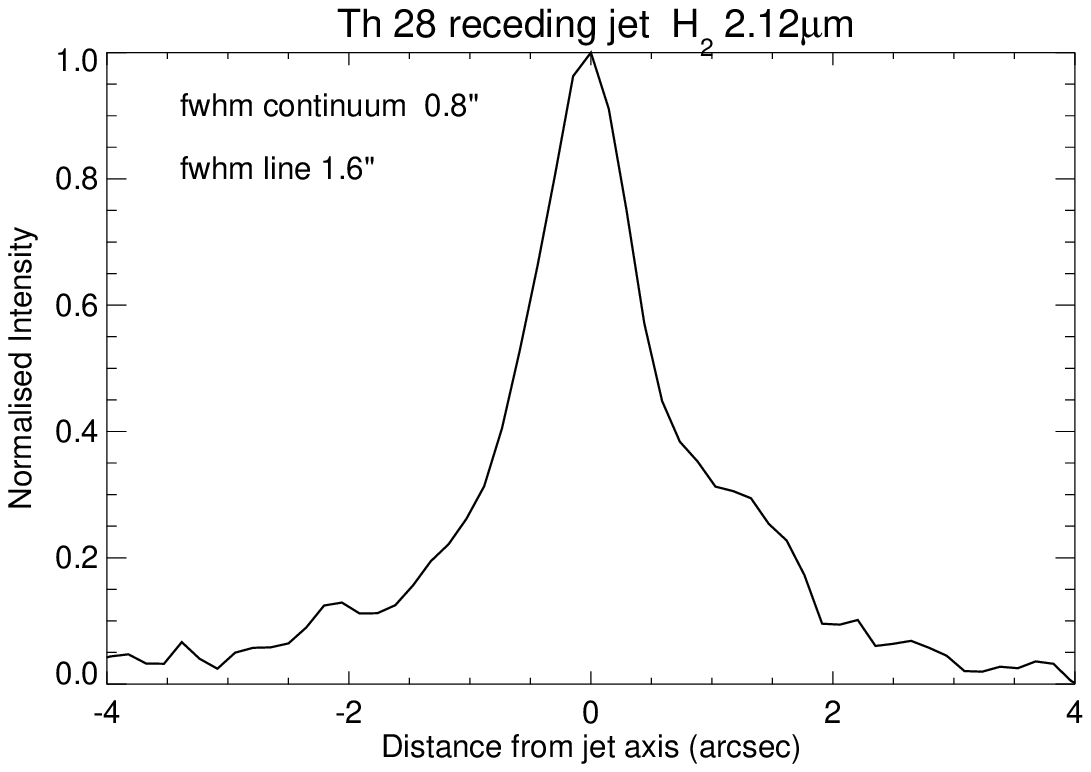}{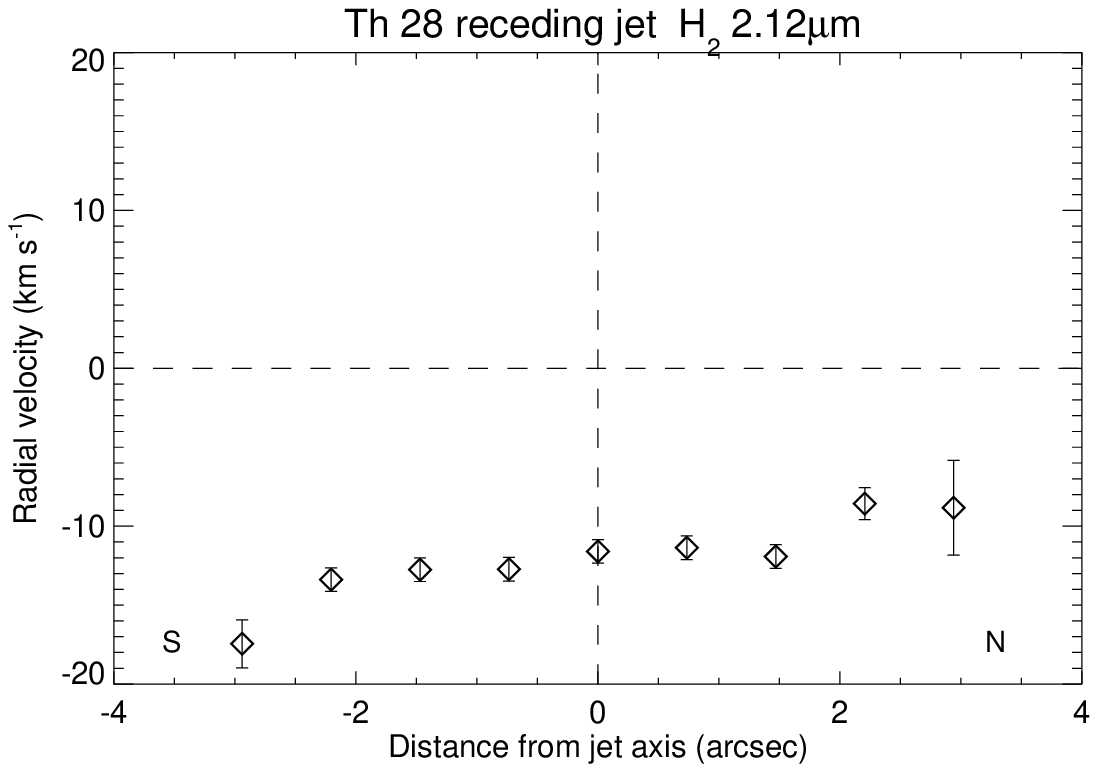}
\caption{Left: PV plot of H$_2$ in the direction of the Th 28 receding jet (via ten linear contours from 3 to 104$\sigma$. ). Middle: the spatial profile of the molecular emission shows a slight asymmetry in emission with respect to the flow axis. Right: transverse velocity profile (spatially binned according to the seeing) of the resolved molecular emission, indicating clearly that the emission is entirely blueshifted even though the slit is positioned over the receding jet lobe. 
\label{h2_perp}}
\end{center}
\end{figure*}

Finally, we consider molecular hydrogen detected in the perpendicular slit spectrum. 
Figure\,\ref{h2_perp} shows a PV diagram, along with the spatial profile and 
the velocity profile obtained by Gaussian fitting. 
As we saw in the parallel-slit spectra, the characteristics of the H$_2$ emission
are {\em unlike} the ionic lines. It is broad enough to be spatially resolved, 
it exhibits a narrow spectral profile extending over 45\,km\,s$^{-1}$, and it
supports the unexpected finding of the parallel-slit data that the emission traces a blue-shifted
gas (-12\,km\,s$^{-1}$) in the region of the receding jet. 
From these data, the H$_2$ gas detected around the base of the bipolar jet is entirely blueshifted to about -10\,km\,s$^{-1}$ in both the parallel and perpendicular slit data regardless of whether the slit was positioned over the approaching or receding jet. It appears that only the blue lobe is emitting in H$_2$, and that we are detecting scattered light from the blue lobe when observing in the direction of the red lobe. This points to an asymmetry in extinction and/or excitation between the two lobes. 
Since the perpendicular slit emission is spatially resolved, we may extract a velocity profile across the gas to determine if there are possible signs of rotation. Figure\,\ref{h2_perp}, right panel, shows a slight systematic gradient in radial velocity, which might give a borderline indication of rotation in the H$_2$ gas. However, the sense of implied rotation is opposite to that measured for optical atomic jet lines resolved with {\it HST}/STIS which presented skews in the profile of 10-25\,km\,s$^{-1}$ for the same target \cite{Coffey04}. 
However, given the nature of this emission, any kinematic asymmetry is more likely caused by environmental factors, rather than intrinsically related to the jet launching mechanism. 

\section{Discussion} 
\label{discussion} 

\subsection{Multi-wavelength jet picture}

These near-IR spectra complement {\it HST}/STIS optical and near-UV observations for the same target, conducted with the same slit position and orientation (proposal ID 9435; 9807). All three datasets agree that atomic emission is fainter for the approaching jet than for the receding jet (\citealp{Coffey04}; 2007), highlighting the fact that the former is more embedded. 

Examining the combined topology, we see that the width of the receding jet was determined to be 0$\farcs$3 and 0$\farcs$4 for optical [\ion{O}{1}] and [\ion{S}{2}] lines respectively \citealp{Coffey04}, while in the near-UV \ion{Mg}{2} lines it was 0$\farcs$1--0$\farcs$2 for both jet lobes \citealp{Coffey07}. Meanwhile, the near-IR atomic lines are not spatially resolved with a seeing of  0$\farcs$7--0$\farcs$8, and the molecular gas presents an outflow width of 1$\farcs$6. This multi-wavelength picture illustrates that the jet constitutes several layers of nested material with the near-UV emitting region at the centre, the optical and near-IR emitting material further out and finally a molecular layer. Indeed, this structure is common among jets (e.g., the jet from DG Tau, see \cite{Coffey07} and \cite{Takami04}). 

Velocities are consistent across all wavelengths regimes. We have seen that [\ion{Fe}{2}] travels at about +30\,km\,s$^{-1}$ in the red lobe and -50\,km\,s$^{-1}$ in the blue lobe. This asymmetry is in excellent agreement with the asymmetry of +30 and -60\,km\,s$^{-1}$ reported for optical emission lines \citep{Coffey04}. Meanwhile, the H$\alpha$ emission is traveling with peak velocities of -127\,km\,s$^{-1}$ and +62\,km\,s$^{-1}$ for the blue and red lobes respectively. (Recall that velocity measurements may be under-estimated, depending on the contribution of scattered light in each emission line.) Such velocity asymmetries between the lobes of a bipolar jet are common, and have been reported in many other cases, e.g., the RW Aur jet \cite{Woitas02} and the LkH$\alpha$233 jet \cite{Melnikov08}. These asymmetries are still not fully explained, but recent estimates of the corresponding mass outflow rates, which transpire to be similar in the two lobes, appear to indicate that the asymmetry is due to environmental conditions on the two sides of the system rather than to an intrinsic asymmetry of the accelerating engine \cite{Melnikov09}. 

The H$_2$ gas which we detect around the base of the bipolar jet is entirely blueshifted to about -10\,km\,s$^{-1}$ in both the parallel and perpendicular slit data regardless of whether the slit was positioned over the approaching or receding jet. It appears that only the blue lobe is emitting in H$_2$, and that we are detecting scattered light from the blue lobe when observing in the direction of the red lobe. This points to an asymmetry in extinction and/or excitation between the two lobes. Also, the receding jet is more luminous in atomic lines. This situation may arise because the approaching jet is more embedded, or because the receding jet is somehow intrinsically more ionised.  However the near-UV \ion{Mg}{2} doublet line ratio is one in the blue lobe (compared to two in the red lobe), which implies a higher column density of \ion{Mg}{2}. This optical thickness in the blue lobe indicates that it is more embedded \cite{Coffey07}, and thus supports an asymmetric extinction. 

\subsection{Electron density and mass flux}

\begin{figure}
\begin{center}
\epsscale{1.0}
\plotone{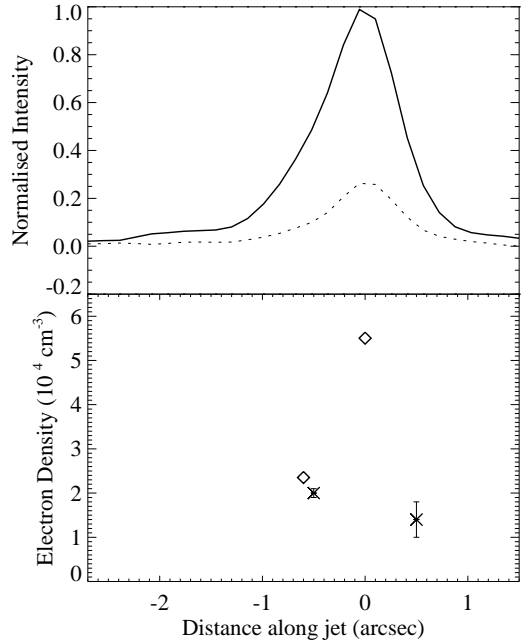}
\caption{Top: Spatial profile of the normalised flux from [Fe II] 1.64 and 1.60$\micron$ emission (solid and dashed lines respectively) from parallel slit observations. Bottom: Electron density derived from the [Fe II] 1.64/1.60 line ratio as a function distance from the source, using parallel and perpendicular slit datasets (diamonds and crosses respectively). For the parallel slit, the flux was spatially binned to match the seeing. Here, the electron density could not be derived for the blue lobe (positive distances along the jet) due to insufficient intensity, unlike the perpendicular slit data which takes in flux from the full jet width. 
\label{ne}}
\end{center}
\end{figure}

The ratio of near infrared iron lines allows us to derive the density in the region, which is usually high enough to saturate the commonly used [\ion{S}{2}] doublet at optical wavelengths. Figure\,\ref{ne} gives the electron density, $n_e$, derived from [\ion{Fe}{2}]1.64/1.60$\mu$m ratio, assuming a temperature 10$^4$\,K. The upper panel gives the normalised intensity of the parallel slit observations, and the lower panel gives electron densities derived from parallel slit fluxes which were binned to match the seeing. Over-plotted are electron densities derived from the perpendicular slit data for the blue and red lobe. Measurements along the jet show that close to the star the electron density saturates the [\ion{Fe}{2}]1.64/1.60$\mu$m ratio, indicating a lower limit of 6\,10$^{4}$\,cm$^{-3}$. However, this value rapidly falls off within the first arcsecond. At a distance from the star of 0$\farcs$5, the electron density is already down to 2\,($\pm$0.1)\,10$^{4}$\,cm$^{-3}$ in the red lobe, as measured in both parallel and perpendicular slit data. Within errors, both lobes can be said to have the same density, although a density asymmetry may exist given the errorbars for the blue lobe i.e., $n_e$=1.4\,($\pm0.4)$\,10$^{4}$\,cm$^{-3}$. Meanwhile, the electron density profile {\em across} the jet is unresolved, and so could not be compared to the density asymmetry derived from the [\ion{S}{2}] doublet in the optical range from {\it HST}/STIS spectra \cite{Coffey08}. The latter report a typical electron density for the red lobe at 0$\farcs$3 from the star of 1\,10$^{4}$\,cm$^{-3}$, although this does not take into account that one side of the jet shows a region which saturates the [\ion{S}{2}] doublet (i.e., lower limit on electron density of 2\,10$^{4}$\,cm$^{-3}$). Finally, the higher electron density closer to the star derived from the iron lines is consistent with the fact that we observe the higher excitation \ion{H}{1} lines of Pa$\beta$ and Br$\gamma$ in this region, as high densities are required to populate the upper levels of these permitted transitions. 

Mass flux estimates are of significant importance in order to access, for example, the drive of molecular flows, their contribution to angular momentum extraction, and the impact of jets and outflows on their environment. Furthermore, accurate estimates of the mass flux rate close to the ejection point of the jet have important implications for the underlying mass loading at the jet base. We determine the mass flux rate, $\dot{M}_{\rm jet}$, following method B of \cite{Nisini05} and \cite{Podio06}. The method derives $\dot{M}_{\rm jet}$ from the observed luminosity, $L(line)$, of
optically thin lines which is proportional to the mass of the emitting gas: 
$\dot{M}_{\rm jet}$ = $\mu$\,$m_{H}$\,($n_{H}$\,$V$)\,$v_{t}$/$l_{t}$, with 
$n_H\,V = L(line)\,\left(h\,\nu\,A_{i}\,f_{i}\,\frac{X^i}{X}\,\frac{X}{H}\right)^{-1}$,
where $\mu$=1.41 is the mean atomic weight, $m_H$ is the proton mass, $n_H$ 
the  hydrogen density, $V$ is the volume filled by the emitting gas, $v_{t}$ and $l_{t}$ are the
tangential velocity and length of the knot, $A_{i}$ and $f_{i}$ are the radiative rate
and the upper level population relative to the considered transition, and 
$\frac{X^i}{X}$ and $\frac{X}{H}$ are the ionisation fraction and the relative
abundance of the considered species. 
This method is affected by uncertainties in absolute calibrations, extinction, and
distance. On the other hand,  it does implicitly take into account the volume filling factor. 
Alternatively, $\dot{M}_{\rm jet}$ may be derived from estimates of the total density and
jet radius. Here, the jet section is assumed uniformly filled at a given density (see Method A of \cite{Podio06}), which may result in an upper limit to the real mass flux. 

To estimate $\dot{M}_{\rm jet}$ at the base of the jet from Th 28, we consider the 
[\ion{Fe}{2}] 1.64 $\mu$m line flux observed using the perpendicular slit. 
Since the red lobe is the brightest, we can measure line flux with high signal-to-noise and thus 
obtain an accurate estimate of electron density. 
We have no estimate of the visual extinction toward Th 28, so we assume A$_{V}$=0,
i.e., we do not correct the derived luminosity for possible reddening effects. 
Finally, we assume a temperature, T, of 10$^4$\,K to compute the upper level population ($f_{i}$), we assume iron
to be completely ionised ($\frac{X^i}{X}$=1) and we assume a solar abundance of iron
($\frac{X}{H}$=2.82 10$^{-5}$ \cite{Asplund05}). Based on these assumption, we find a value of $\sim$2 10$^{-9}$ and 1 10$^{-9}$$M_{\odot}$\,yr$^{-1}$ for the mass flux rate of the receding and approaching jet respectively. The red lobe yields twice the mass flux of the blue lobe. This is due to the lower {\em apparent} flux of the blue lobe ($\sim$20\% of the red lobe flux), which is only slightly compensated by the fact that the velocity is higher in the blue lobe (i.e., twice the red lobe velocity). The mass flux asymmetry however does not account for an asymmetric extinction between the two lobes. 

Our values are lower than the typical mass flux rate of $\sim$10$^{-8}$ $M_{\odot}$yr$^{-1}$ reported for T Tauri stars \cite{Hartigan95}. The outcome is a lower limit on the mass flux rate for two reasons. First, we do not have an estimate for the extinction and, second, the iron gas phase abundance may be depleted with respect to the solar abundance (\cite{Podio06}; 2008). The value reported from optical lines \cite{Coffey08} is 1.2~10$^{-8}$ $M_{\odot}$yr$^{-1}$, calculated by using the total density and jet radius, i.e., method A. This order of magnitude difference in mass flux rate between the two methods (i.e., relying on the luminosity as opposed to electron density and ionisation determinations) is the typical difference found in previous studies which compare the two method. Note that for this jet target, the flux of the optical lines is roughly two orders of magnitude higher than the [\ion{Fe}{2}] 1.64$\micron$ flux. This can be due to iron depletion and/or intrinsically fainter emission in the iron lines with respect to the optical lines in this target. 

\subsection{\ion{H}{1} Origin}

There is ongoing debate as to whether the origin of hydrogen line emission close to the star is accretion and/or ejection related (Section\,\ref{introduction}). Our results demonstrate that a component of both Pa$\beta$ and Br$\gamma$ emission is undoubtedly coming from the outflow. This is evident from the clear spatial extension seen in the parallel slit data, and from the velocities found in these spatially extended wings. The finding is interesting given that these are high-excitation lines. Both lines present similar velocities to the [\ion{Fe}{2}] lines which we know to trace the jet, and are further confirmed to originate in jet by the detection of the similarly high excitation [\ion{Fe}{2}] 2.13 $\micron$ line. This is a highly excited iron line, which has been detected in jet knots in conjunction with Br$\gamma$ (see e.g., HH 34 in \cite{GarciaLopez08}). It is a further indication that Br$\gamma$ originates in the jet. The velocities and fluxes of these \ion{H}{1} lines measured close to the star (Table\,\ref{onedspec_table}) provide important constraints on models. The proportion of flux originating in the outflow could not be determined because these lines arise in the higher density jet region close to the star which suffers terms of seeing limitations and scattered light. Finally, the importance of identifying Pa$\beta$ and Br$\gamma$ as originating in the outflow lies in the fact that they are permitted transitions in the near-IR. Permitted transitions emit in high density regions where forbidden lines are quenched, and near-IR lines are less affected by extinction than optical lines.  Hence, the near-IR lines Pa$\beta$ and Br$\gamma$ become very important probes of the embedded central engine of protostars. The development of high resolution interferometric instruments in the infrared, such as VLTI/AMBER, means it is important to identify in advance useful probes of the central engine. 

\section{Conclusions}
\label{conclusion}

Our study provides the first information on the near-IR spectrum of the microjet from Th 28, revealing emission in [\ion{Fe}{2}], Pa$\beta$, Br$\gamma$ and H$_2$. This near-IR dataset complements the optical and near-UV data taken for the same target with {\it HST}/STIS. The Th\,28 outflow now constitutes the only bipolar jet system to date for which we will have a complete spectroscopic dataset sampling all the coaxial layers of the jet, and for both jet lobes. 

The molecular H$_2$ emission was detected surrounding the base of both jet lobes, but its blueshifted velocity of -10\,km\,s$^{-1}$ indicates that only the blue shifted lobe is emitting while light is scattered in the direction of the red lobe, highlighting an asymmetric extinction and/or excitation between the two lobes. Outflow radial velocities in ionic lines are in agreement across all wavelengths though the velocities, and the velocity asymmetry between the lobes. The {\it HST}/STIS H$\alpha$ emission, observed at 0$\farcs$3 from the star, presents radial velocities of -127\,km\,s$^{-1}$ and +62\,km\,s$^{-1}$ for the blue and red lobe respectively. Unlike optical and near-UV results, there is no indication of jet rotation in [\ion{Fe}{2}] lines because of insufficient angular resolution to resolve the jet width. The iron ratio was found to saturate close to the star implying a lower limit on the electron densities of 6$\times$10$^4$\,cm$^{3}$. However, the high density falls off rapidly to 2\,10$^4$\,cm$^{3}$ within the first arcsecond along each flow, and leads to a value of 2 and 1~10$^{-9}$ $M_{\odot}$yr$^{-1}$ for the mass flux rate of the receding and approaching jet calculated from iron line {\it apparent} luminosity. This is an order of magnitude lower than that derived using optical lines, likely due to differences inherent in the method and possible depletion of iron in the gas phase. Finally, a velocity analysis, along with the detection of the high excitation [\ion{Fe}{2}] 2.13 $\micron$ line, reveals that the jet is traced in Pa$\beta$ and Br$\gamma$ within the first 150\,AU from the star. These higher excitation hydrogen lines are usually attributed to accretion activity, but our observations clearly demonstrate there is a component which traces outflow.  These observations, though seeing limited, have good spectral resolution and thus will complement the high angular resolution of VLTI/AMBER interferometric observations when sensitivities are sufficient to observe T Tauri jet emission. 

\vspace {0.2in}
{\bf Acknowledgements} 
\vspace {0.1in}
\newline
The present work was supported in part by the European Community's Marie Curie Actions - Human Resource and Mobility within the JETSET (Jet Simulations, Experiments and Theory) network, under contract MRTN-CT-2004-005592. D.C. and L.P. are supported by the Irish Research Council for Science, Engineering and Technology (IRCSET). This work is based on observations obtained with ESO VLT/ISAAC, Paranal, Chile, within the observing program 077.C-0818A. This work is also based on observations made with the NASA/ESA {\it Hubble Space Telescope}, obtained at the Space Telescope Science Institute, which is operated by the Association of Universities for Research in Astronomy, Inc., under NASA contract NAS5-26555.


\end{document}